\documentclass[a4paper,11pt]{article}
\pdfoutput=1 % if your are submitting a pdflatex (i.e. if you have
\usepackage{jheppub} % for details on the use of the package, please
\usepackage[T1]{fontenc} % if needed

\usepackage{graphicx}
\usepackage{multirow}
\usepackage{amsmath}
\usepackage{bm}
\usepackage{slashed}
\usepackage{epsfig}
\usepackage{amsfonts}
\usepackage{color}
\usepackage{epstopdf}
\usepackage{enumitem}
\allowdisplaybreaks
\newcommand{\ud}{\mathrm{d}}
\newcommand{\uv}{\mathrm{UV}}
\newcommand{\ir}{\mathrm{IR}}

\newcommand{\gev}{\mathrm{~GeV}}

\newcommand{\msbar}{\overline{\mathrm{MS}}}

\newcommand{\jpsi}{{J/\psi}}

\newcommand{\state}[4]{{^{#1}\hspace{-0.6mm}#2_{#3}^{[#4]}}}

\newcommand\CSaPa{\state{1}{P}{1}{1}}
\newcommand\CScSa{\state{3}{S}{1}{1}}
\newcommand\CScPz{\state{3}{P}{0}{1}}
\newcommand\CScPa{\state{3}{P}{1}{1}}
\newcommand\CScPb{\state{3}{P}{2}{1}}
\newcommand\CScPj{\state{3}{P}{J}{1}}

\newcommand\COcSa{\state{3}{S}{1}{8}}

\newcommand\COcPj{\state{3}{P}{J}{8}}

\newcommand\mo{{\mathcal O}}

\newcommand{\LDME}[2]{\langle\mo^{#1}(#2)\rangle}

\newcommand{\LDMEn}[1]{\langle\mo^{#1}_n\rangle}

\newcommand\mohn{\LDMEn{H}}

\def\gev{\mathrm{~GeV}}

\def\co{{\cal O}}

\title{\boldmath Gluon fragmentation into ${^{3}\hspace{-0.6mm}P_{J}^{[1,8]}}$ quark pair and\\ test of NRQCD factorization at two-loop level}

\author[a,b]{Peng Zhang,}
\author[a]{Ce Meng}
\author[a,b,c]{Yan-Qing Ma,}
\author[a,b,c]{and Kuang-Ta Chao}

\affiliation[a]{School of Physics and State Key Laboratory of Nuclear Physics and Technology, Peking University,\\Beijing 100871, China}
\affiliation[b]{Center for High Energy Physics, Peking University,\\Beijing 100871, China}
\affiliation[c]{Collaborative Innovation Center of Quantum Matter,\\Beijing 100871, China}

\emailAdd{p.zhang@pku.edu.cn}
\emailAdd{mengce75@pku.edu.cn}
\emailAdd{yqma@pku.edu.cn}
\emailAdd{ktchao@pku.edu.cn}

\abstract{The next-to-leading order (NLO) ($\mathcal{O}(\alpha_s^3)$) corrections for gluon fragmentation functions to a heavy quark-antiquark pair in $\state{3}{P}{J}{1,8}$ states are calculated within the NRQCD factorization. We use integration-by-parts reduction and differential equations to semi-analytically calculate fragmentation functions in full-QCD, and find that infrared divergences can be absorbed by the NRQCD long distance matrix elements. Thus, the NRQCD factorization conjecture is verified at two-loop level via a physical process, which is free of artificial ultraviolet divergences. Through matching procedure, infrared-safe short distance coefficients and $\mathcal{O}(\alpha_s^2)$ perturbative NRQCD matrix elements $\LDME{\state{3}{P}{J}{1/8}}{\COcSa}_{\mathrm{NLO}}$ are obtained simultaneously. The NLO short distance coefficients are found to have significant corrections comparing with the LO ones. }

\keywords{Perturbative calculations, Factorization, Heavy quarkonia}%Use showqkeys class option if keyword
\begin{document}

\maketitle

\flushbottom

\section{Introduction}

Heavy quarkonium provides an ideal physical system to explore the strong interaction, as the heavy quarkonium production contains both perturbative and non-perturbative effects in QCD. Non-relativistic QCD (NRQCD) factorization \cite{Bodwin:1994jh} is the most widely used theory to explain quarkonium production so far.
When transverse momentum $p_T$ of the produced quarkonium is large, fragmentation mechanism dominates and factorization is easier to hold, because  long-distance interactions between quarkonium and initial-state particles are suppressed then.

Inclusive production differential cross section of a specific hadron $H$ at high $p_T$ can be calculated in collinear factorization \cite{Collins:1989gx},
\begin{equation}
	\ud \sigma _{A+B\rightarrow H(p_T)+X} = \sum_i \ud \hat{\sigma} _{A+B \rightarrow i(p_T /z)+X'} \otimes D_{i\to H}(z,\mu) + \co (1/p_T^2) \, ,
\end{equation}
where $i$ sums over all quarks and gluons, $z$ is the light-cone momentum fraction carried by $H$ with respect to the parent parton $i$, and $A$ and $B$ are colliding particles whose effect should be further factorized into partons if they are hadrons.
$d \hat{\sigma} _{A+B \rightarrow i(p_T /z)+X}$ are perturbatively calculable hard parts, while $D_{i\to H}(z,\mu)$ are non-perturbative but universal fragmentation functions (FFs) describing the probability of parton hadronizing to $H$ with momentum fraction $z$. For heavy quarkonium production, an important $\co (1/p_T^2)$ contributions are double parton FFs which can also be factorized \cite{Kang:2011mg,Kang:2014tta,Kang:2014pya,Fleming:2012wy,Fleming:2013qu}.
In FFs, there is a collinear factorization scale $\mu$ dependence, which cancels with the similar dependence in hard parts perturbatively order by order, leaving physical differential cross section independent of the scale. The evolution of single parton FFs with respect to $\mu$ are controlled by the Dokshitzer-Gribov-Lipatov-Altarelli-Parisi (DGLAP) evolution equation \cite{Gribov:1972ri,Altarelli:1977zs,Dokshitzer:1977sg}, and similar evolution equations for double parton FFs are calculated in \cite{Kang:2014tta}. With these evolution equations, the only unknown information for FFs are their values at any chosen factorization scale $\mu=\mu_f$.

When $\mu_f$ is close to the quarkonium mass $m_H$, it is natural to calculate FFs via NRQCD factorization. For single parton FFs that will be considered in this paper, we have
\begin{equation} \label{eq:ff}
	D_{i\to H}(z,\mu_f) = \sum_n d_{i\to Q\bar{Q}(n)} (z,\mu_f) \langle \bar{\mo}^{H}_n\rangle \, ,
\end{equation}
where $d_{i\to Q\bar{Q}(n)}$ represent the perturbative calculable short-distance coefficients (SDCs) to produce a heavy quark-antiquark pair $Q\bar Q$ with quantum number $n$, and $\langle \bar{\mo}^{H}_n\rangle$ are normalized long-distance matrix elements (LDMEs) \footnote{ $\langle \bar{\mo}^{H}_n\rangle$ can be related to the original definition of NRQCD LDME $\mohn$ \cite{Bodwin:1994jh} by the following rules. They are the same if $n$ is color-octet, and $\langle \bar{\mo}^{H}_n\rangle=\mohn/(2N_c)$ if $n$ is color-singlet.}. The quantum number is usually expressed in spectroscopic notation $n=\state{{2S+1}}{L}{J}{c}$, with $c=1, 8$ respectively for color-singlet state and color-octet state. According to velocity scaling rule \cite{Bodwin:1994jh},  $\langle \bar{\mo}^{H}_n\rangle$ is usually suppressed if $L$ is too large. Therefore, the most important states for phenomenological purpose are $S$-wave and $P$-wave states. Because LDMEs are supposed to be process independent, they can be determined by fitting experimental data, while SDCs can be calculated perturbatively through the matching procedure.

All the SDCs for single parton FFs to both $S$-wave and $P$-wave states up to $\mathcal{O}(\alpha_s^2)$ are available \cite{ Braaten:1993mp,Braaten:1993rw,Cho:1994gb,Braaten:1994kd,Beneke:1995yb,Ma:1995vi,Braaten:1996rp,Braaten:2000pc,Hao:2009fa,Jia:2012qx,Bodwin:2014bia} (see  \cite{Ma:2013yla,Ma:2015yka} for a summary and comparison). At $\alpha_s^3$ order, the leading order (LO) SDCs of $g \rightarrow Q\bar{Q}(\CScSa)+X$ and $g \rightarrow Q\bar{Q}(\CSaPa)+X$ are calculated separately in Refs. \cite{Zhang:2017xoj,Braaten:1993rw,Braaten:1995cj,Bodwin:2003wh,Bodwin:2012xc} and Ref. \cite{Sun:2018yam}, and the next-to-leading order (NLO) SDCs of $g\to Q\bar{Q}(\state{1}{S}{0}{1,8})+X$ have also been calculated in Refs. \cite{Zhang:2018mlo,Artoisenet:2018dbs,Feng:2018ulg} recently. To perform phenomenological analysis at a consistent precision, such as that for $\jpsi$ or $\chi_{cJ}$ production, at this order, one also needs the NLO SDCs of $g \rightarrow Q\bar{Q}(\state{3}{P}{J}{1,8})+X$ and the next-to-next-to-leading order (NNLO) SDCs of $g \rightarrow Q\bar{Q}(\COcSa)+X$. In this paper, we will focus on the former.

Another motivation to calculate the NLO FFs for processes $g \rightarrow Q\bar{Q}(\state{3}{P}{J}{1,8})+X$ is to test the NRQCD factorization conjecture at two-loop level with a physical process. The test at two-loop  level exists in literature \cite{Nayak:2005rw,Nayak:2005rt,Nayak:2006fm,Bodwin:2019bpf} but within the framework of  eikonal approximation  for infrared (IR) divergences. The eikonal approximation introduces artificial ultraviolet (UV) divergences which makes the extraction of IR divergences  beyond one-loop level highly nontrivial. Because we do not use eikonal approximation, the extraction of IR divergences in our case is straightforward.
%For the processes $g \rightarrow Q\bar{Q}(\state{3}{P}{J}{1})+X$, the infrared (IR) divergences in the LO full-QCD FFs  are absolutely absorbed in the $\mathcal{O}(\alpha_s)$ perturbative NRQCD matrix element (PME) $\LDME{\CScPj}{\COcSa}_{\mathrm{LO}}$~\cite{Braaten:2000pc,Ma:2013yla}, which verified the factorization conjecture at one-loop level. At two-loop level, the IR divergent part in the NLO full-QCD FFs has been calculated in Ref. \cite{Nayak:2005rt,Bodwin:2019bpf} with the help of eikonal approximation. The total IR divergences can be absorbed into the PMEs at the same order, which proves NRQCD factorization at two-loop level. In this paper, we will perform the two-loop proof more strictly by the full-QCD calculations of the NLO FFs without eikonal approximation.
%Besides, with the NLO FF $g \rightarrow Q\bar{Q}(\COcPj)+X$, it becomes possible to discuss the phenomenology of $\jpsi$ production.

In this paper, we aim to calculate NLO SDCs of $g \rightarrow Q\bar{Q}(\state{3}{P}{J}{1,8})+X$ to high precision using similar methods in our previous paper \cite{Zhang:2018mlo}. The rest of the paper is organized as following. In Sec. \ref{sec:full}, we calculate the process to LO and NLO in full-QCD. The calculation method is very similar to that in Ref. \cite{Zhang:2018mlo}. A small improvement is the computation of operator renormalization shown in Sec. \ref{sec:nlo} and Appendix. \ref{sec:ori}.
We argue that the NRQCD factorization at two-loop level without eikonal approximation is verified in Sec. \ref{sec:test}. The $\mathcal{O}(\alpha_s^2)$ perturbative NRQCD matrix elements (PMEs) $\LDME{\CScPj}{\COcSa}_{\mathrm{NLO}}$ and $\LDME{\COcPj}{\COcSa}_{\mathrm{NLO}}$ are matched, and the corresponding IR-safe SDCs are obtained simultaneously in Sec. \ref{sec:match}. Discussion will be presented in Sec. \ref{sec:sdcdis}.

\section{Full-QCD Calculation} \label{sec:full}

\subsection{Definition}

FF of a gluon to a hadron (quarkonium) is defined by Collins and Soper \cite{Collins:1981uw},
\begin{align}\label{eq:defFF}
\begin{split}
    D_{g \rightarrow H}(z,\mu_0)=
    & \frac{-g_{\mu\nu}z^{D-3}}{2 \pi P_c^{+}(N_{c}^{2}-1)(D-2)} \int_{-\infty}^{+\infty}\mathrm{d}x^{-} e^{-i P_c^{+} x^{-}} \\
    & \times \langle 0 | G_{c}^{+\mu}(0) \mathcal{E}^{\dag}(0,0,\boldsymbol{0}_{\perp})_{cb} \mathcal{P}_{H(P)} \mathcal{E}(0,x^{-},\boldsymbol{0}_{\perp})_{ba} G_{a}^{+\nu}(0,x^{-},\boldsymbol{0}_{\perp}) | 0 \rangle \, ,
\end{split}
\end{align}
where $G^{\mu\nu}$ is the gluon field-strength operator, $P$ and $P_c$ are respectively the momenta of the produced hadron $H$ and the initial-state fragmenting gluon $g$, and $z=P^+/P_c^+$ is the ratio of momenta along the ``+'' direction.
It is convenient to choose the frame in which the hadron has zero transverse momentum, $P = (z P_c^+, M^2/(2 z P_c^+),\boldsymbol{0}_{\perp})$, with $P^2=2P^+P^-=M^2$.
The projection operator $\mathcal{P}_{H(P)}$ is defined by
\begin{equation}\label{eq:projectH}
\mathcal{P}_{H(P)} = \sum_X |H(P)+X \rangle \langle H(P)+X|\,,
\end{equation}
where $X$ sums over all unobserved particles.
The gauge link $\mathcal{E}(x^{-})$ is an eikonal operator that involves a path-ordered exponential of gluon field operators along a light-like path,
\begin{equation}
  \mathcal{E}(0,x^{-},\boldsymbol{0}_{\perp})_{ba}= \mathrm{P} \, \text{exp} \left[+i g_s \int_{x^{-}}^{\infty}\mathrm{d}z^{-} A^{+}(0,z^{-},\boldsymbol{0}_{\perp}) \right]_{ba} \, ,
\end{equation}
where $g_s=\sqrt{4 \pi \alpha_s}$ is the QCD coupling constant and $A^{\mu}(x)$ is the matrix-valued gluon field in the adjoint representation: $[A^{\mu}(x)]_{ac} = i f^{abc} A^{\mu}_{b}(x)$.

Since the SDCs $d$'s in the NRQCD factorization formula (\ref{eq:ff}) are independent of the final state $H$, we can replace $H$ by on-shell $Q\bar{Q}$ states to extract $d_{i\to Q\bar{Q}[n]}$ through the matching procedure [See below (\ref{eq:matchlo}) and (\ref{eq:matchnlo})].

Choosing the factorization scale $\mu_0 \sim 2m_Q$ with $m_Q$ being the heavy quark mass, we can calculate FF for $g\to Q\bar{Q}[n]$ in the full-QCD with the Feynman gauge.
Most of the Feynman rules are the same as those of QCD, while the others related to the eikonal line can be found in Ref.~\cite{Zhang:2018mlo}. Feynman amplitudes are denoted as $\mathcal{M}_{\lambda_{{Q}}\lambda_{\bar{Q}}\lambda_i}(q)$, where $\lambda_{{Q}}$ and $\lambda_{\bar{Q}}$ are respectively spins of produced on-shell heavy quark and heavy antiquark, $\lambda_i$ ($i=1, 2, \dots$) are spins of the initial-state virtual gluon or final-state unobserved light particles, and $q$ is the heavy quark momentum in the rest frame of heavy quark pair.
To project the free $Q\bar{Q}$ pair to the state with spin-triplet and specific color, one can multiply amplitudes by projection operators and obtain
\begin{equation}
	\mathcal{M}^\alpha_{\lambda_i}(q)=\text{Tr}\left[ \Gamma_c \Gamma_\lambda^\alpha \overline{\mathcal{M}}_{\lambda_{{Q}}\lambda_{\bar{Q}}\lambda_i}(q)  \right]\, ,
\end{equation}
where $\overline{\mathcal{M}}_{\lambda_{{Q}}\lambda_{\bar{Q}}\lambda_i}$ denotes amplitude with external heavy-(anti)quark spinors removed and the projection operators are defined as
\begin{align}
\begin{split}
	&\Gamma_{c=1}
	 = \frac{1} {\sqrt{N_{c}}} \,, \\
	&\Gamma_{c=8}
	 = \frac{\sqrt{2} T^{a}} {\sqrt{N_{c}^{2}-1}} \,, \\
	&\Gamma_\lambda^\alpha
	 = \frac{1} {\sqrt{2 E} (E + m_Q)}
		(\slashed{\overline{p}} - m_Q)
		\frac{2 E - \slashed{P}}{4 E}
		\gamma^\alpha
		\frac{2 E + \slashed{P}}{4 E}
		(\slashed{p} - m_Q)	\, ,	
\end{split}
\end{align}
where $p=P/2+q$ and $\overline{p}=P/2-q$ are momenta respectively for $Q$ and $\bar{Q}$,  $m_Q$ is the mass of heavy quark, and $E$ is the energy of $Q$ or $\bar{Q}$ in the rest frame of $Q\bar{Q}$. With the on-shell conditions of $Q$ and $\bar{Q}$, we have
\begin{align}
	P\cdot q =0,  \quad \quad P^2=4E^2=4(m_Q^2-q^2).
\end{align}
To project the amplitude into P-wave, we also need to take the derivation over $q^\beta$,
\begin{equation}
	\mathcal{M}^{\alpha \beta}_{\lambda_i} =
	 \frac{\partial}{\partial q^\beta}
	\mathcal{M}^\alpha_{\lambda_i}(q) |_{q\to 0} \,.
\end{equation}
% \begin{equation}
% 	\mathcal{M}^\beta_{\lambda \lambda_0 \lambda_i}(P,k_{i},m_Q) =
% 	\left| \frac{\partial}{\partial q^\beta}
% 	\mathcal{M}_{\lambda \lambda_0 \lambda_i}(P,q,k_{i},m_Q) \right|_{q\to 0} \,.
% \end{equation}
And by summing over spin and color of initial-state and final-state particles, we get the squared amplitude
\begin{equation}
	\mathcal{A}_J  = \frac{N_{\mathrm{CS}}}{N_J} \overline{\sum}
	P_J^{\alpha \alpha' \beta \beta'} \,
	\mathrm{Re} \left[ \mathcal{M}^{\alpha \beta}_{\lambda_i} \mathcal{M}^{*\alpha' \beta'}_{\lambda_i} \right]\,,
\end{equation}
% \begin{equation}
% 	\mathcal{A}_J ({P},{k}_{i},m_Q) =N_{\mathrm{CS}} \overline{\sum}
% 	I_J^{\beta \beta'} (P) \,
% 	\mathrm{Re} \left[ \mathcal{M}^\beta_{\lambda \lambda_0 \lambda_i}(P,k_{i},m_Q) \mathcal{M}^{*\beta'}_{\lambda \lambda_0 \lambda_i}(P,k_{i},m_Q) \right]\,,
% \end{equation}
where $N_{\mathrm{CS}}=\frac{z^{D-2}}{(N_{c}^{2}-1)(D-2)} $ with $D=4-2\epsilon$ is the space-time dimension.
% and $z$ is the ``$+$'' momentum fraction of the initial virtual gluon carried by the hadron.
For different $J$, $N_J$ and $P_J^{\alpha \alpha' \beta \beta'}$ are respectively
\begin{equation}
	N_J = \left\{
	\begin{aligned}
 		& 1 \,, & \text{for } J=0 \\
 		& \frac{1}{2}(D-1)(D-2) \,, & \text{for }    J=1 \\
 		& \frac{1}{2}(D+1)(D-2)\,, & \text{for } J=2 \\
 		& (D-1)^2 \,, & \text{for } \sum_J
 	\end{aligned}
	\right. \,,
\end{equation}
and
\begin{equation} \label{eq:project}
	P_J^{\alpha \alpha' \beta \beta'} = \left\{
	\begin{aligned}
 		& \frac{1}{D-1} I^{\alpha \beta} I^{\alpha' \beta'} \,, & \text{for } J=0 \\
 		& \frac{1}{2}\left( I^{\alpha \alpha'} I^{\beta \beta'}-I^{\alpha \beta'} I^{\alpha' \beta} \right) \,, & \text{for }    J=1 \\
 		& \frac{1}{2}\left( I^{\alpha \alpha'} I^{\beta \beta'}-I^{\alpha \beta'} I^{\alpha' \beta} \right) - \frac{1}{D-1} I^{\alpha \beta} I^{\alpha' \beta'} \,, & \text{for } J=2 \\
 		& I^{\alpha \alpha'} I^{\beta \beta'} \,, & \text{for } \sum_J
 	\end{aligned}
	\right. \,,
\end{equation}
where
\begin{equation}
	I^{\alpha\beta} = -g^{\alpha \beta}
			+ \frac{P^{\alpha} P^{\beta}} {P^{2}} \, .
\end{equation}
Then, fragmentation function in the full-QCD can be denoted as
\begin{equation} \label{eq:FFint}
	D[g\to Q\bar{Q} (\state{3}{P}{J}{1/8}])] = \int \ud \Phi \mathcal{A}_J \, .
\end{equation}
Here, $\ud \Phi$ is the final-state phase space,
\begin{equation}
    \mathrm{d} \Phi =
  	\frac {P^{+}}{z^{2} S}
  	\delta \left( \frac{1-z}{z} P^{+} - \sum_i k_{i}^{+} \right)
    \prod_{i} \frac{\mathrm{d}^{D} k_i}{(2\pi)^{D-1}}
        \delta _+ (k_i^{2})
         \,,
\end{equation}
where $S$ is the symmetry factor for final-state particles and $k_i$'s are momenta of final-state light particles.
In the rest of the paper, we will take the case of $g\to Q\bar{Q} (\CScPz)+X$ as an explicit example to explain our calculations.

\subsection{LO FFs} \label{sec:lo}
\begin{figure}[htb!]
 \begin{center}
 \includegraphics[width=120pt]{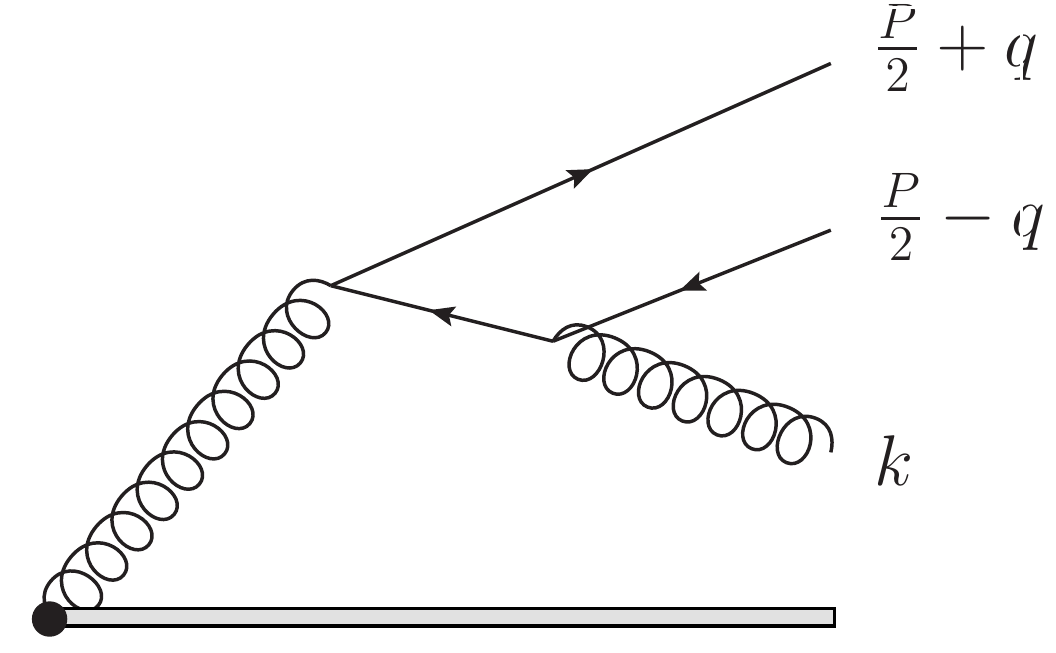}
  \caption{One of the two Feynman diagrams of gluon fragmenting into $Q\bar{Q}[\state{3}{P}{0}{1}]$ at LO in $\alpha_s$. Another diagram can be obtained by permuting the heavy quark and anti-quark.  \label{fig:lofd}}
 \end{center}
\end{figure}

The Feynman diagrams of gluon fragmenting into $Q\bar{Q}[\state{3}{P}{0}{1}]$ at the LO in $\alpha_s$ are shown in Figure  \ref{fig:lofd}.
These diagrams can be easily calculated, and the obtained FF can be expressed as
\begin{equation}\label{eq:lo}
	D_{\mathrm{LO}}[g\to Q\bar{Q} (\CScPz)] = \frac{N_{\mathrm{LO}}}{1-\epsilon}
 		\left[ (1-z)^{-\epsilon} f_1(z) + (1-z)^{-1-2\epsilon} f_2(z) \right] \, ,
\end{equation}
where
\begin{equation}\label{eq:factorlo}
	N_{\mathrm{LO}} = \frac{ \, \alpha_s^2 }{(D-1) N_c m_Q^5 }
 		\left( \frac{\pi \mu_r^2}{m_Q^2} \right)^{\epsilon} \Gamma(1+\epsilon)
\end{equation}
and
\begin{align}
\begin{split}
	f_1(z) & = \frac{3(5-3z)}{\epsilon }+\frac{3}{2} (3 z-4)-\frac{9 z  }{2}\epsilon \,,\\
	f_2(z) & = \frac{3(1-z) (3z-5)}{\epsilon }+\frac{13 z^3-42 z^2+15 z+18}{3}  +2 \left(2 z^2-2 z-1\right) z \epsilon +\frac{2 z^3 \epsilon ^2}{3}
			% & \phantom{= \frac{1}{(3-2\epsilon)(1-\epsilon)} \Bigg(}
  \,.
\end{split}
\end{align}
With the following decomposition
\begin{equation}
	(1-z)^{-1-2\epsilon} = -\frac{\delta(1-z)}{2 \epsilon }+\frac{1}{ [1-z]_+}-2 \epsilon  \left[ \frac{\ln(1-z)  }{1-z}\right]_+ + \mathcal{O} (\epsilon^2) \,,
\end{equation}
the FF can be expressed in the limit of $\epsilon \to 0$ as
\begin{align}\label{eq:loexp}
\begin{split}
	D_{\mathrm{LO}}[g\to Q\bar{Q} (\CScPz)] = & N_{\mathrm{LO}}
	\Big( -\frac{2}{3 \epsilon} \delta(1-z)+\frac{1}{3} \delta(1-z)+\frac{4}{3} \frac{z}{[1-z]_+} \\
	& \phantom{ N_{\mathrm{LO}}\Big( }
	+\frac{1}{6} (85-26 z) z+3 (5-3 z) \ln(1-z) \Big) \,.
\end{split}
\end{align}
It is clear that there is an IR divergence in the full-QCD result.
This divergence can be absorbed by the color-octet PME, which will be explained in Sec. \ref{sec:match}.

\subsection{NLO FFs} \label{sec:nlo}

\begin{figure}[htb!]
 \begin{center}
 \includegraphics[width=\textwidth]{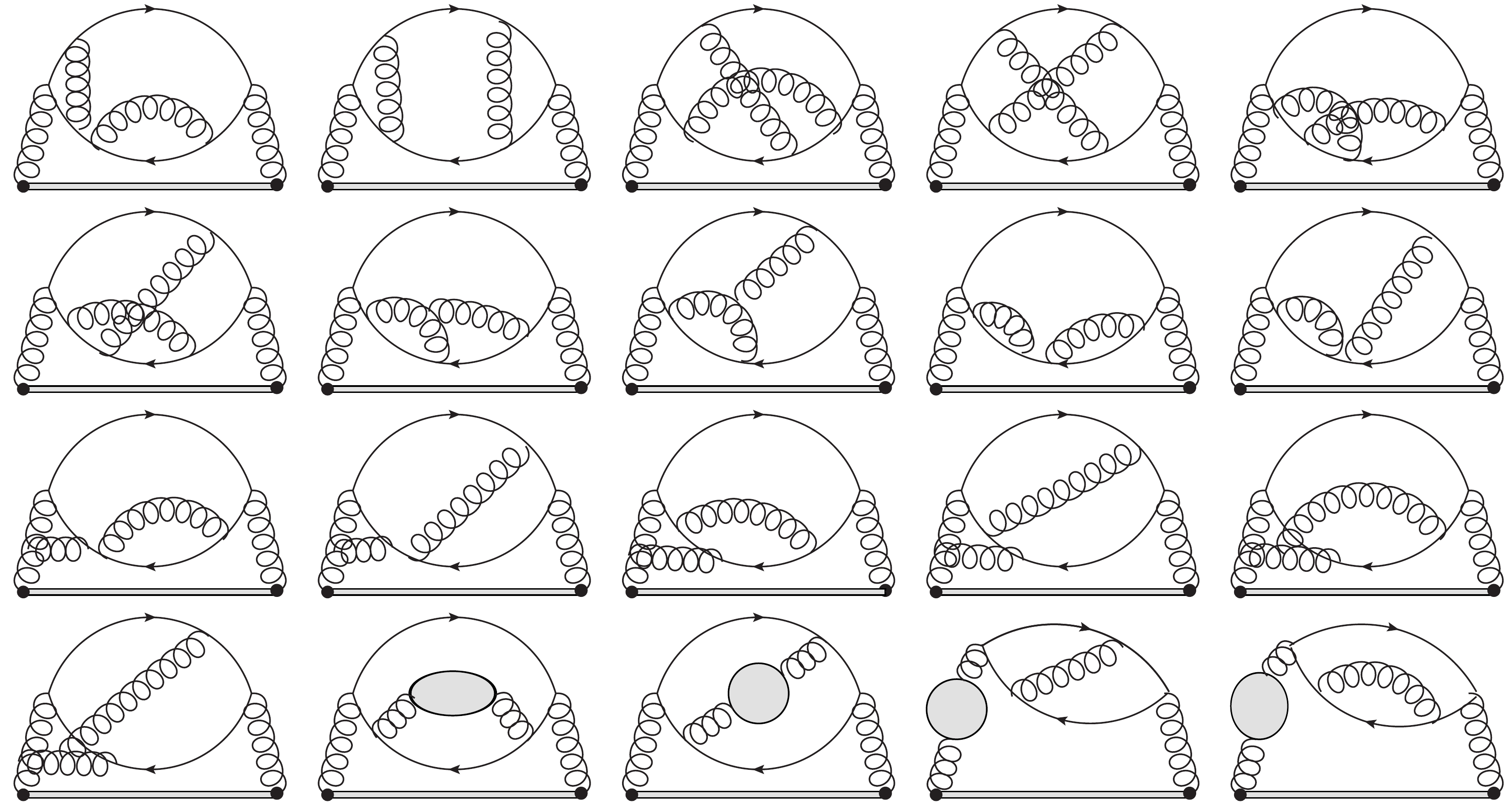}
  \caption{Typical Feynman diagrams without gluon eikonal vertex for $g\to Q\bar{Q}(\state{3}{P}{0}{1})+X$. The other diagrams can be obtained by permuting the heavy quark and anti-quark. The real and virtual diagrams can be obtained by cutting at different positions.  \label{fig:nlofd}}
 \end{center}
\end{figure}

\begin{figure}[htb!]
 \begin{center}
 \includegraphics[width=0.8\textwidth]{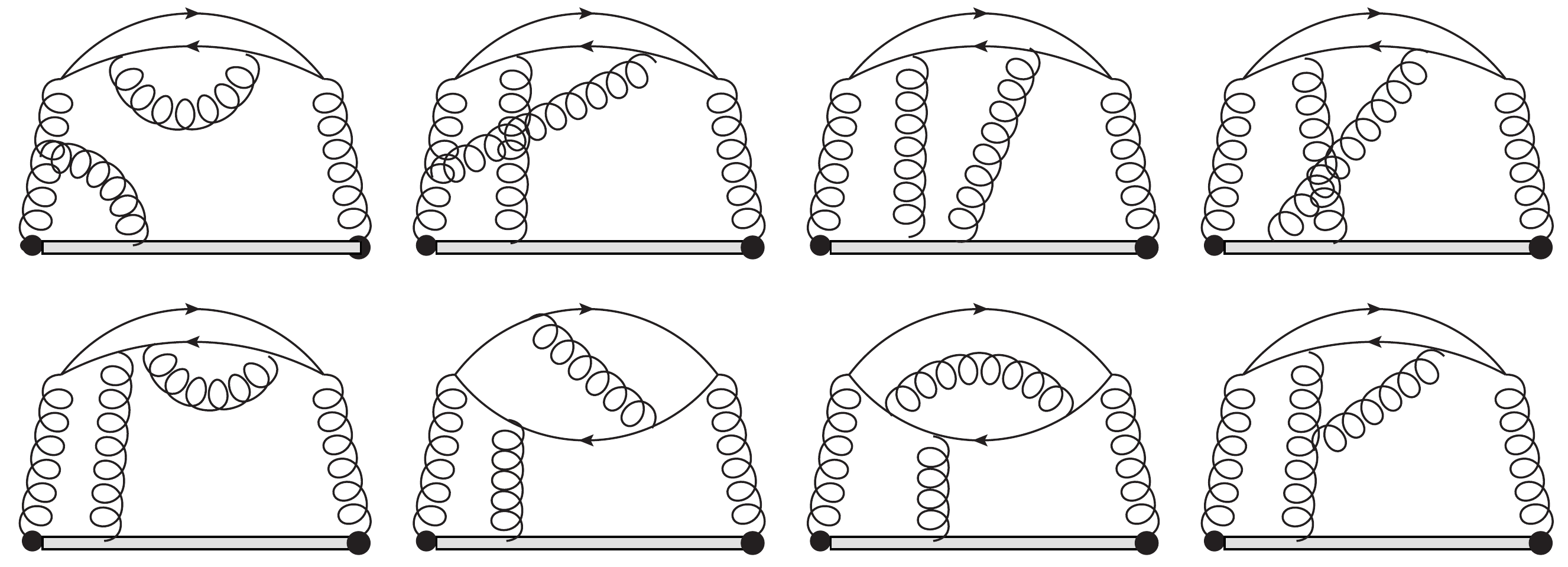}
  \caption{Typical Feynman diagrams with gluon eikonal vertex for $g\to Q\bar{Q}(\state{3}{P}{0}{1})+X$. The other diagrams can be obtained by permuting the heavy quark and anti-quark. The real and virtual diagrams can be obtained by cutting at different positions.  \label{fig:nloekfd}}
 \end{center}
\end{figure}

\begin{figure}[htb!]
 \begin{center}
 \includegraphics[width=0.5\textwidth]{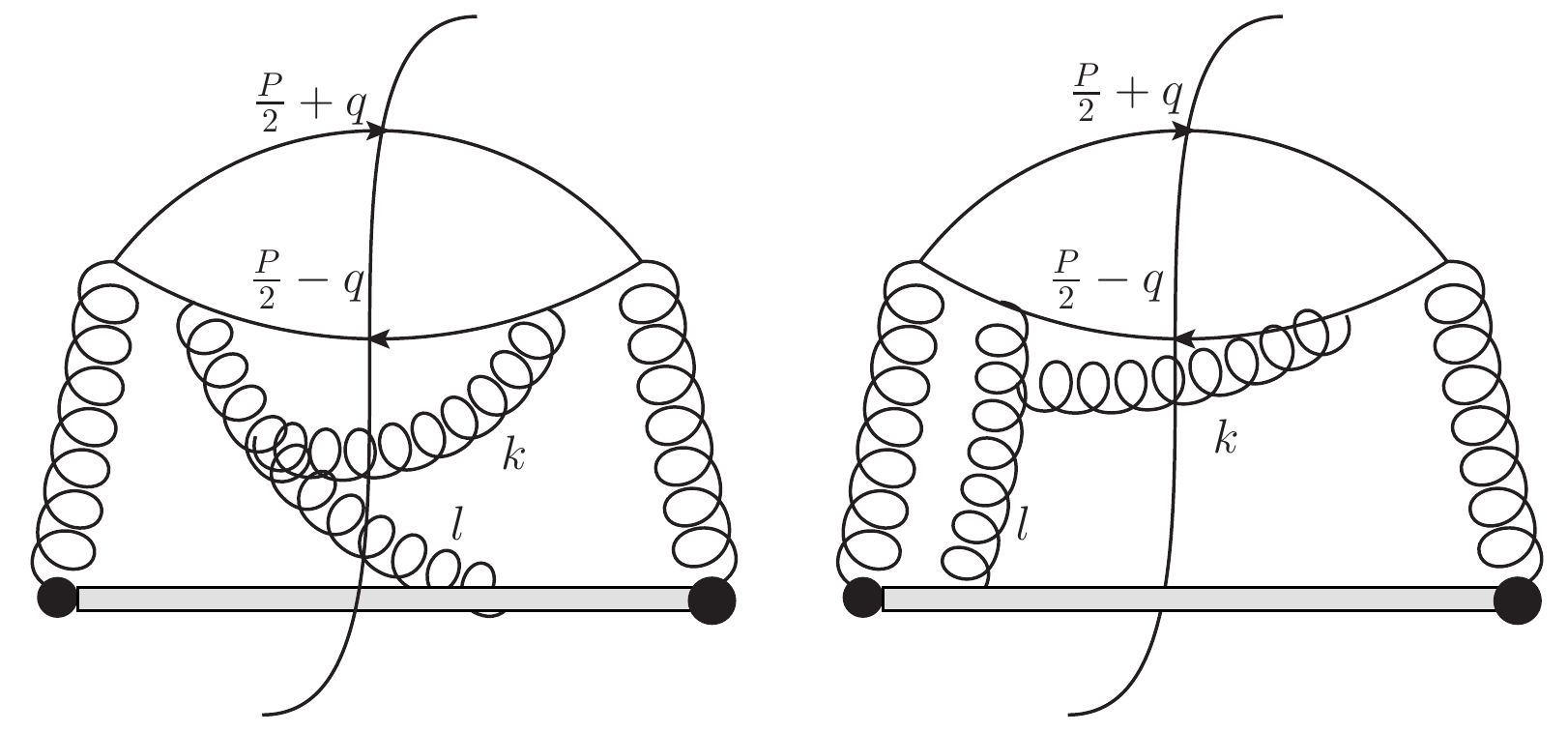}
  \caption{ Real and virtual cut diagrams by cutting at different positions for the last diagrams in Figure \ref{fig:nloekfd}.  \label{fig:cutfd}}
 \end{center}
\end{figure}

The NLO full-QCD calculations of this process resemble closely to the NLO calculations of $g\to Q\bar{Q}(\state{1}{S}{0}{1,8})+X$ in Ref. \cite{Zhang:2018mlo}.
The relevant Feynman diagrams are shown in Figure \ref{fig:nlofd} and Figure \ref{fig:nloekfd}. In our calculation, we need to cut these diagrams at some specific positions. In Figure \ref{fig:cutfd}, we give an example to show how to obtain cut diagrams for the real and  the virtual corrections.

The calculation is almost the same as that in Ref. \cite{Zhang:2018mlo}. Firstly we generate the amplitude from the Feynman diagrams for real or virtual corrections. Then reduce them into linear combinations of master integrals (MIs). There are 95 MIs for real corrections and 66 MIs for virtual corrections.
Fortunately, these MIs are the same as those in Ref. \cite{Zhang:2018mlo}, which can be calculated by using differential equations (DEs) method \cite{Kotikov:1990kg,Bern:1992em,Remiddi:1997ny,Gehrmann:1999as,Henn:2013pwa, Lee:2014ioa,Adams:2017tga,Caffo:2008aw,Czakon:2008zk,Mueller:2015lrx,Lee:2017qql,Liu:2017jxz,Liu:2020kpc}.
We can estimate values of these MIs in regions $0\sim1/4$, $1/4\sim3/4$ and $3/4\sim1$ respectively by the asymptotic expansions of them at $z=0,1/2$ and $1$.
Finally, the real and virtual corrections can be expressed by asymptotic expansions in different regions as
\begin{equation}\label{eq:rvexp}
	\sum_{s} \sum_{i=0}^{n_{s}} (z-z_0)^{s} \ln^{i}(z-z_0) \sum_{j=0}^{\infty} I^{ s \, i \, j}(\epsilon) (z-z_0)^j \, ,
\end{equation}
where $s$ are linear functions of $\epsilon$, $n_s$ are integers determined by $s$, and coefficients $I^{ s \, i \,j }(\epsilon)$ are functions of $\epsilon$, which can be computed to sufficient high order with high precision.

Renormalization has some differences from that in Ref. \cite{Zhang:2018mlo}, because $1/\epsilon$ pole already presents at LO FF.
Firstly, the on-shell renormalization constants of heavy quark field and mass ($\delta_2$ and $\delta_m$) should be expanded at least to $\mathcal{O}(\epsilon)$.
To simplify the procedure, we use expression with exact $\epsilon$ dependence,
\begin{equation}
	\delta_i = \frac{\alpha_s}{\pi} \Gamma(1+\epsilon)
				\left( \frac{\pi \mu_r^2}{m_Q^2} \right)^{\epsilon}
				\hat{\delta} _i \, ,
\end{equation}
with
\begin{align}\label{eq:zren}
\begin{split}
	\hat{\delta}_2 & = -\frac{C_F}{4^{1-\epsilon}}
						\left( \frac{1}{\epsilon_{\uv}} \frac{1+2 \epsilon}{1-2 \epsilon } +\frac{2}{\epsilon_{\ir}} \right) \,, \\
	\hat{\delta}_m & = -\frac{C_F}{4^{1-\epsilon}}
						\frac{1}{\epsilon_{\uv}} \frac{3-2 \epsilon }{1-2 \epsilon } \,,
\end{split}
\end{align}
where $C_F=(N_c^2-1)/(2N_c) $.

Apart from the calculations of the QCD counter terms,  some tricks are needed in the calculation of operator renormalization.
In the $\msbar$ scheme, the operator counter term is given by
\begin{equation}\label{eq:operator}
	D_{\mathrm{NLO} }^{\mathrm{Operator}} (z) =
		-\frac{\alpha_s}{2\pi}
		\frac{\Gamma(1+\epsilon)}{\epsilon}
		\left( \frac{4 \pi \mu_r^2}{\mu_0^2} \right)^{\epsilon}
		\int_z^1 \frac{\ud y}{y} P_{gg} (y)
			D_{\mathrm{LO}} \left(\frac{z}{y} \right) \, ,
\end{equation}
where $\mu_0$ is the collinear factorization scale, and the Altarelli-Parisi splitting function $P_{gg} (z)$ is
\begin{equation}\label{eq:splitting}
	P_{gg} (z) = b_0 \, \delta (1-z) +
				 2 N_c \left( \frac{z}{(1-z)_+} + \frac{1-z}{z} + z(1-z) \right) \, ,
\end{equation}
where $b_0 = (11N_c-2n_f)/6$.
If one takes the Eq.~\eqref{eq:loexp} as LO FF $D_{\mathrm{LO}}(z)$, there will be two plus functions convolved in the integral in (\ref{eq:operator}), which is difficult to define and calculate.
However, we can take Eq.~\eqref{eq:lo} as the input form of the $D_{\mathrm{LO}}(z)$ to overcome this difficulty, with corresponding integrated results given in Appendix \ref{sec:ori}.
In this way, operator renormalization results can be easily obtained in each region, which are convenient to be added to the real and virtual corrections.

\subsection{Verification of the NRQCD factorization at two-loop level}
\label{sec:test}

%As shown in Eq.~\eqref{eq:ff}, FF can be expressed as summations of different LDMEs multiplied by SDCs. If NRQCD factorization is valid, all IR divergences should be absorbed into the LDMEs, leaving SDCs IR-safe.

In the full-QCD calculations, the FF can be expressed by the phase space integral of amplitude squared as shown in Eq.~\eqref{eq:FFint}. Before taking the derivative over $q$, the amplitude squared at ${\cal{O}}(q^2)$ can be expressed as
\begin{equation}\label{eq:ampqn}
	\mathcal{A} (q,q') =  (q\cdot q')\mathcal{A}_1 + \frac{(q\cdot n) (q'\cdot n)}{(P\cdot n) ^2} \mathcal{A}_2 \,,
\end{equation}
where $n$ is the light-like vector that defines the direction of the eikonal line of fragmentation function. Because NRQCD LDMEs should be independent of the direction of the eikonal line of fragmentation function, and all remained IR divergences in full-QCD calculation must be absorbed into the LDMEs if NRQCD factorization is valid, there should be no IR divergences in $\mathcal{A}_2$. On the other hand, if one can show that there is no IR divergence in $\mathcal{A}_2$, then all IR divergences in full-QCD FF can be absorbed into the NRQCD LDMEs in principle, and the NRQCD factorization can be verified at two-loop level.

Two-loop verification of NRQCD factorization was also done in Refs. \cite{Nayak:2005rw,Nayak:2005rt,Nayak:2006fm,Bodwin:2019bpf}, in which eikonal approximation was used to simplify the calculation. The eikonal approximation however introduced artificial UV divergences that makes the extraction of IR divergences highly nontrivial.

In this subsection, we verify the NRQCD factorization without applying eikonal approximation. We take the process $g\to Q\bar{Q} (\CScPj)+X$ with $J$ summed as an example.
%In the full-QCD calculation before, we take the projection opertator $P_J^{\alpha \alpha' \beta \beta'}$ as
%\begin{equation}
%	 P_J^{\alpha \alpha' \beta \beta'} = I^{\alpha \alpha'} I^{\beta \beta'} \,
%\end{equation}
%with which the $P$-wave amplitude squared in Eq.~\eqref{eq:ampqn} can be projected as
%\begin{equation}
%	P_J^{\alpha \alpha' \beta \beta'} \, \frac{\partial}{\partial q^\beta \, \partial q'^{\beta'}} \mathcal{A} (q,q')
%	= I^{\alpha \alpha'}\left(- (D-1) \mathcal{A}_1 + \frac{1}{P^2} \mathcal{A}_2 \right) \,.
%\end{equation}
%Thus, there will be a mix between $\mathcal{A}_1$ and $\mathcal{A}_2$.
To extract the part $\mathcal{A}_2$ only, we use the following projection operator
%can simply change the projection operator $P_J^{\alpha \alpha' \beta \beta'}$ to
\begin{equation}
	P_{qn}^{\alpha \alpha' \beta \beta'} =  I^{\alpha \alpha'}
	\left( (D-2)I^{\beta \beta'} -(D-1)d^{\beta \beta'} \right)  \,,
\end{equation}
where
\begin{equation}
	d^{\beta \beta'} = -g^{\beta \beta'} + \frac{P^{\beta} n^{ \beta'} + P^{ \beta'} n^{\beta}} {P\cdot n}
			- \frac{P^2 n^{\beta} n^{ \beta'}} {(P\cdot n)^{2}} \,,
\end{equation}
and we get
\begin{equation}
	P_{qn}^{\alpha \alpha' \beta \beta'} \, \frac{\partial}{\partial q^\beta \, \partial q'^{\beta'}} \mathcal{A} (q,q')
	= I^{\alpha \alpha'} (D-2)\frac{1}{P^2} \mathcal{A}_2  \,.
\end{equation}
Performing the calculations similar to those introduced in Sec.~\ref{sec:full}, we find that all divergences are canceled out after adding real corrections, virtual corrections as well as UV renormalization contributions. As a result, IR divergences in $\mathcal{A} (q,q')$ are free of the gauge line vector $n$. Thus the NRQCD factorization is verified to hold at two-loop level, which is consistent with the observation in Refs. \cite{Nayak:2005rw,Nayak:2005rt,Nayak:2006fm,Bodwin:2019bpf}.

%Real and virtual corrections can be reduced to MIs, which can be expressed by asymptotic expansions at $z=0,1/2,1$ as shown in Eq.~\eqref{eq:rvexp}.
%The counter term and operator renormalization can be calculated analytically and also be expanded at $z=0,1/2,1$.
%With the IBP reduction, UV and IR divergences will be mixed. However the UV divergence will be canceled by summing over the real, virtual, counter term and operator renormalization. Then the IR divergence is left, which is also canceled for each asymptotic expansions through calculation.
%Thus NRQCD factorization is proved to this order.

\section{Matching procedure and the SDCs}

%\subsection{NRQCD Matching}
\label{sec:match}

As  SDCs are independent of specifics of the hadronic states,  we can match the $\mathcal{O}(\alpha_s^2)$ PMEs from the divergences in full-QCD FFs and obtain the IR-safe SDCs simultaneously, thanks to the fact that NRQCD factorization holds at this order.

For $g\to Q\bar{Q}(\state{3}{P}{J}{1/8} )$ process, the LO FF can be expressed by
\begin{align}\label{eq:matchlo}
\begin{split}
	D_{\mathrm{LO}}[g\to Q\bar{Q} (\state{3}{P}{J}{1/8})] =
	& \phantom{+} d_{\mathrm{LO}}[g\to Q\bar{Q} (\state{3}{P}{J}{1/8})] \LDME{\state{3}{P}{J}{1/8}}{\state{3}{P}{J}{1/8}}_{\mathrm{LO}} \\
	& + d_{\mathrm{LO}}[g\to Q\bar{Q} (\COcSa)] \LDME{\state{3}{P}{J}{1/8}}{\COcSa}_{\mathrm{LO}} \,,
\end{split}
\end{align}
where $\LDME{\CScPj}{\CScPj}_{\mathrm{LO}}=\LDME{\COcPj}{\COcPj}_{\mathrm{LO}} =1$ is normalized. The LO SDC of gluon to $\COcSa$ in (\ref{eq:matchlo}) has been calculated in Ref. \cite{Braaten:1996rp,Braaten:2000pc}, which is given in $D-$dimension by
\begin{equation}
	d_{\mathrm{LO}}[g\to Q\bar{Q} (\COcSa)] =
	\frac{\pi  \alpha_s \delta (1-z)}{ (D-1) (N_c^2-1) m_Q^3} \,,
\end{equation}
and the $\mathcal{O}(\alpha_s)$ PMEs in the $\msbar$ subtraction scheme can be expressed as
\begin{align}
	\LDME{\CScPj}{\COcSa}_{\mathrm{LO}} &=
	- C_F \frac{4\alpha_s}{3 \pi m_Q^2}
	\frac{\Gamma(1+\epsilon)}{\epsilon_\ir}
	\left( \frac{4 \pi \mu_r^2}{\mu_\Lambda^2} \right)^{\epsilon} \,,\\
	\LDME{\COcPj}{\COcSa}_{\mathrm{LO}} &=
	- B_F \frac{4\alpha_s}{3 \pi m_Q^2}
	\frac{\Gamma(1+\epsilon)}{\epsilon_\ir}
	\left( \frac{4 \pi \mu_r^2}{\mu_\Lambda^2} \right)^{\epsilon} \,,
\end{align}
where $B_F=(N_c^2-4)/(4N_c) $ and $\mu_\Lambda$ is the NRQCD factorization scale.
With full-QCD FF in Eq.~\eqref{eq:loexp}, the $D-$dimension SDC $d_{\mathrm{LO}}[g\to Q\bar{Q} (\CScPj)]$ can be matched as
\begin{align}\label{eq:sdclomu}
\begin{split}
	d_{\mathrm{LO}}[g\to Q\bar{Q} (\CScPz)] =&\frac{ \, \alpha_s^2 \Gamma(1+\epsilon)}{(D-1) N_c m_Q^5 }
 		\left( \frac{\pi \mu_r^2}{m_Q^2} \right)^{\epsilon} \times \Bigg(  \left(\frac{1}{3} - \frac{2}{3} \ln \left(\frac{\mu_\Lambda ^2}{4 m_Q^2} \right) \right) \delta(1-z)  \\
	&+\frac{4}{3} \frac{z}{[1-z]_+}
	 +\frac{1}{6} (85-26 z) z+3 (5-3 z) \ln(1-z) \Bigg) \,,
\end{split}
\end{align}
which equals to the finite part of the full-QCD FF in Eq.~\eqref{eq:loexp} if factorization scale is chosen as $\mu_\Lambda = 2m_Q$.
We find that the IR divergence in full-QCD FF $D_{\mathrm{LO}}[g\to Q\bar{Q} (\CScPz)]$ is absorbed into the $\mathcal{O}(\alpha_s)$ PME $\LDME{\CScPj}{\COcSa}_{\mathrm{LO}}$, and the SDCs are IR-safe up to LO.
This result is consistent with that in Ref. \cite{Ma:2013yla}.

Similarly, $D_{\mathrm{NLO}}[g\to Q\bar{Q} (\state{3}{P}{J}{1/8})]$ can be expressed in the NRQCD factorization as
\begin{align}\label{eq:matchnlo}
\begin{split}
	D_{\mathrm{NLO}}[g\to Q\bar{Q} (\state{3}{P}{J}{1/8})] = \phantom{+}
	& d_{\mathrm{NLO}}[g\to Q\bar{Q} (\state{3}{P}{J}{1/8})] \LDME{\state{3}{P}{J}{1/8}}{\state{3}{P}{J}{1/8}}_{\mathrm{LO}} \\
	+ & d_{\mathrm{LO}}[g\to Q\bar{Q} (\state{3}{P}{J}{1/8}))] \LDME{\state{3}{P}{J}{1/8}}{\state{3}{P}{J}{1/8}}_{\mathrm{NLO}} \\
	+ & d_{\mathrm{NLO}}[g\to Q\bar{Q} (\COcSa)] \LDME{\state{3}{P}{J}{1/8}}{\COcSa}_{\mathrm{LO}} \\
	+ & d_{\mathrm{LO}}[g\to Q\bar{Q} (\COcSa)] \LDME{\state{3}{P}{J}{1/8}}{\COcSa}_{\mathrm{NLO}} \,,
\end{split}
\end{align}
where all LO terms are known. For NLO terms, $D_{\mathrm{NLO}}[g\to Q\bar{Q} (\state{3}{P}{J}{1/8})]$ are calculated in this work, $\LDME{\state{3}{P}{J}{1/8}}{\state{3}{P}{J}{1/8}}_{\mathrm{NLO}}$ are well-known, which have only Coulomb divergences  and vanish in dimensional regularization with $\msbar$ subtraction scheme, and $d_{\mathrm{NLO}}[g\to Q\bar{Q} (\COcSa)]$ has been calculated in Refs. \cite{Braaten:2000pc,Ma:2013yla}, whose $D-$dimension form is given by
\begin{align}
\begin{split}
	d_{\mathrm{NLO}}[g\to Q\bar{Q} (\COcSa)] = &
	\frac{ \alpha_s^2 \Gamma(1+\epsilon)}{ 4(D-1) C_F m_Q^3}
	\left( \frac{\pi \mu_r^2}{m_Q^2} \right)^{\epsilon}
	\Bigg[ A(\mu_r) \delta(1-z) + \frac{1}{N_c} P_{gg} (z) \times  \\
	& \left( \ln \left(\frac{\mu_0 ^2}{4 m_Q^2} \right) -1 \right)+\frac{2(1-z)}{z} - \frac{4(1-z+z^2)^2}{2} \left[\frac{\ln(1-z)}{1-z} \right]_+
	 \Bigg] \,,
\end{split}
\end{align}
with
\begin{equation}
	A(\mu_r) = \frac{b_0}{N_c} \left( \ln \left(\frac{\mu_r ^2}{4 m_Q^2} \right) +\frac{13}{3} \right)
	+ \frac{4}{N_c^2} -\frac{\pi^2}{3}+ \frac{16}{3}\ln 2 \,.
\end{equation}
Therefore, SDCs $d_{\mathrm{NLO}}[g\to Q\bar{Q} (\state{3}{P}{J}{1/8})]$ and PMEs $\LDME{\state{3}{P}{J}{1/8}}{\COcSa}_{\mathrm{NLO}}$ can be extracted from Eq.~\eqref{eq:matchnlo} thanks to the fact that the former one is finite and the later one is IR-divergent defined in $\msbar$ subtraction scheme.

%By chosen overall factors
%\begin{equation}\label{eq:factornlo}
%	N_{\mathrm{NLO}} = \frac{ \, \alpha_s^3 }{(D-1) \pi N_c m_Q^5 }
% 		\left( \frac{\pi \mu_r^2}{m_Q^2} \right)^{2\epsilon} \Gamma^2(1+\epsilon) \,,
%\end{equation}
%for the color-singlet process and
%\begin{equation}\label{eq:factornlo8}
%	N^{[8]}_{\mathrm{NLO}} = \frac{N_c^2-4}{2(N_c^2-1)} \frac{ \, \alpha_s^3 }{(D-1) \pi N_c m_Q^5 }
% 		\left( \frac{\pi \mu_r^2}{m_Q^2} \right)^{2\epsilon} \Gamma^2(1+\epsilon) \,,
%\end{equation}
%for the color-octet process.

The $\mathcal{O}(\alpha_s^2)$ PMEs $\LDME{\state{3}{P}{J}{1/8}}{\COcSa}_{\mathrm{NLO}}$ are obtained numerically with high precision.
Using PSLQ algorithm \cite{ferguson1999analysis,Bailey:1999nv}, analytical expressions are obtained as
\begin{align}
	\LDME{\CScPj}{\COcSa}_{\mathrm{NLO}} &=
	- C_F \frac{\alpha_s^2}{27 \pi^2 m_Q^2}
	\left( \frac{4 \pi \mu_r^2}{\mu_\Lambda^2} \right)^{2\epsilon}
	\Gamma(1+\epsilon)^2
	\left( \frac{9b_0}{\epsilon^2}
	+\frac{5n_f+(12\pi^2-47)N_c}{\epsilon} \right)   \,,\\
		\LDME{\COcPj}{\COcSa}_{\mathrm{NLO}} &=
	- B_F \frac{\alpha_s^2}{27 \pi^2 m_Q^2}
	\left( \frac{4 \pi \mu_r^2}{\mu_\Lambda^2} \right)^{2\epsilon}
	\Gamma(1+\epsilon)^2
	\left( \frac{9b_0}{\epsilon^2}
	+\frac{5n_f+(3\pi^2-47)N_c}{\epsilon} \right)   \,.
\end{align}

Setting factorization scales $\mu_0=\mu_\Lambda=2m_Q$, the NLO SDC can be expressed as
\begin{align}\label{eq:sdcnlo}
\begin{split}
	d_{\mathrm{NLO}}[g\to Q\bar{Q} (\CScPz)] =
	\frac{ \alpha_s^3 }{3 \pi N_c m_Q^5 }
	\Bigg(
	& p_\delta \delta (1-z)
	+ \sum_{i=0}^2 p_i \left[\frac{\ln^i(1-z)}{1-z}\right]_+
	+ p(z) \\
	& +\ln \left(\frac{\mu_r^2}{4m_Q^2} \right) b_0 \, d_{\mathrm{LO}}[g\to Q\bar{Q} (\CScPz)]
	\Bigg) \,,
\end{split}
\end{align}
where the coefficients $p_\delta,p_i$ can be fitted with the PSLQ algorithm from high-precision numerical values as following
\begin{align}
\begin{split}
	p_\delta & = \frac{95-24 \pi ^2}{162}  n_f
		+\frac{756 \zeta (3)+66 \pi ^2-1141-576 \ln ^2 2-1032 \ln 2}{324} N_c
		- \frac{2(4 \ln 2-1)}{3} \frac{1}{N_c} \,,\\
	p_0 & = -n_f
		+\frac{-12 \pi ^2+119+64 \ln 2 }{18}  N_c
		+ \frac{8}{3} \frac{1}{N_c} \,,\\
	p_1 & = \frac{4}{9}  n_f
		-\frac{64 }{9}  N_c \,,\\
	p_2 & = -\frac{16 }{3}  N_c  \,,
\end{split}
\end{align}
and $p(z)$ can be expressed as a piecewise function
\begin{equation}
  	p(z) =
  	\left\{
 	\begin{aligned}
 		&\frac{63-8 \pi ^2}{36 z} N_c + \sum_{i=0}^{2} \sum_{j=0}^{\infty} \ln^i z \, (2z)^j \left(A_{ij}^{f} \, n_f + A_{ij} \, N_c +  \frac{A_{ij}^{N}}{N_c} \right) \,, & \text{for } 0<z<\frac{1}{4} \\
 		&\sum_{j=0}^{\infty} (2z-1)^j \left(B_{j}^{f} \, n_f + B_{j}\, N_c + \frac{B_{j}^{N}}{N_c} \right) \,, & \text{for }    \frac{1}{4} \le z \le \frac{3}{4} \\
 		&\sum_{i=0}^{3} \sum_{j=0}^{\infty} \ln^i (1-z) \, (2-2z)^j \, \left(C_{ij}^{f} \, n_f + C_{ij} \, N_c + \frac{C_{ij}^{N}}{N_c} \right) \,. & \text{for } \frac{3}{4}<z<1
 	\end{aligned}
 	\right. \,.
\end{equation}
The coefficients $A_{ij}^{k},B_{j}^{k},C_{ij}^{k}$ can be evaluated numerically with high precision. In the attached ancillary file, we give these coefficients calculated up to $j=500$ to obtain the results with $140$-digit precision at any value of $z$.

The SDCs of $g\to Q\bar{Q} (\CScPj)+X$ with $J=1,2$ and $g\to Q\bar{Q} (\COcPj)+X$ with sum $J$ are calculated similarly, with results given as attachments in the ancillary file.

\section{Numerical Results and Discussion} \label{sec:sdcdis}

\begin{figure}
\begin{minipage}{0.5\textwidth}
  \centerline{$g\to b\bar{b} (\CScPz)+X $}
  \centerline{\includegraphics[width=1\textwidth]{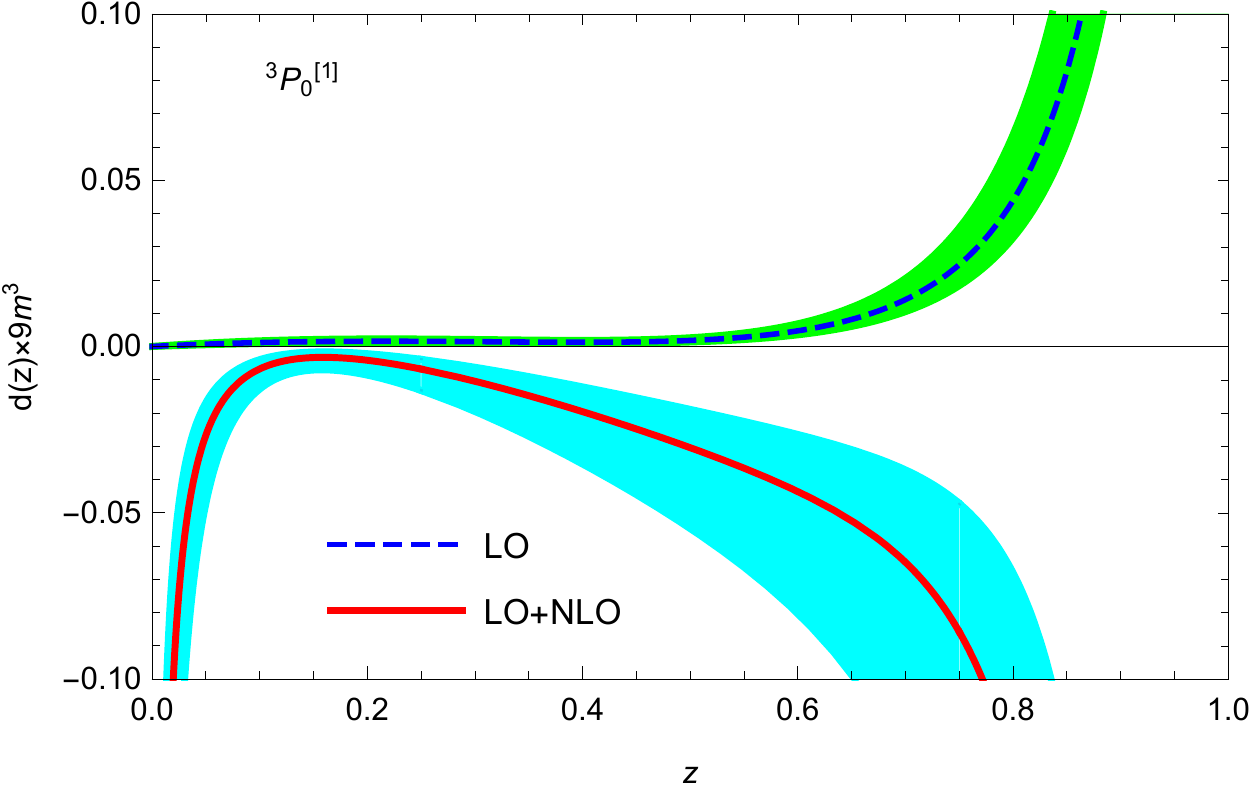}}
\end{minipage}
\hfill
\begin{minipage}{0.5\textwidth}
  \centerline{$g\to b\bar{b} (\CScPa)+X $}
  \centerline{\includegraphics[width=1\textwidth]{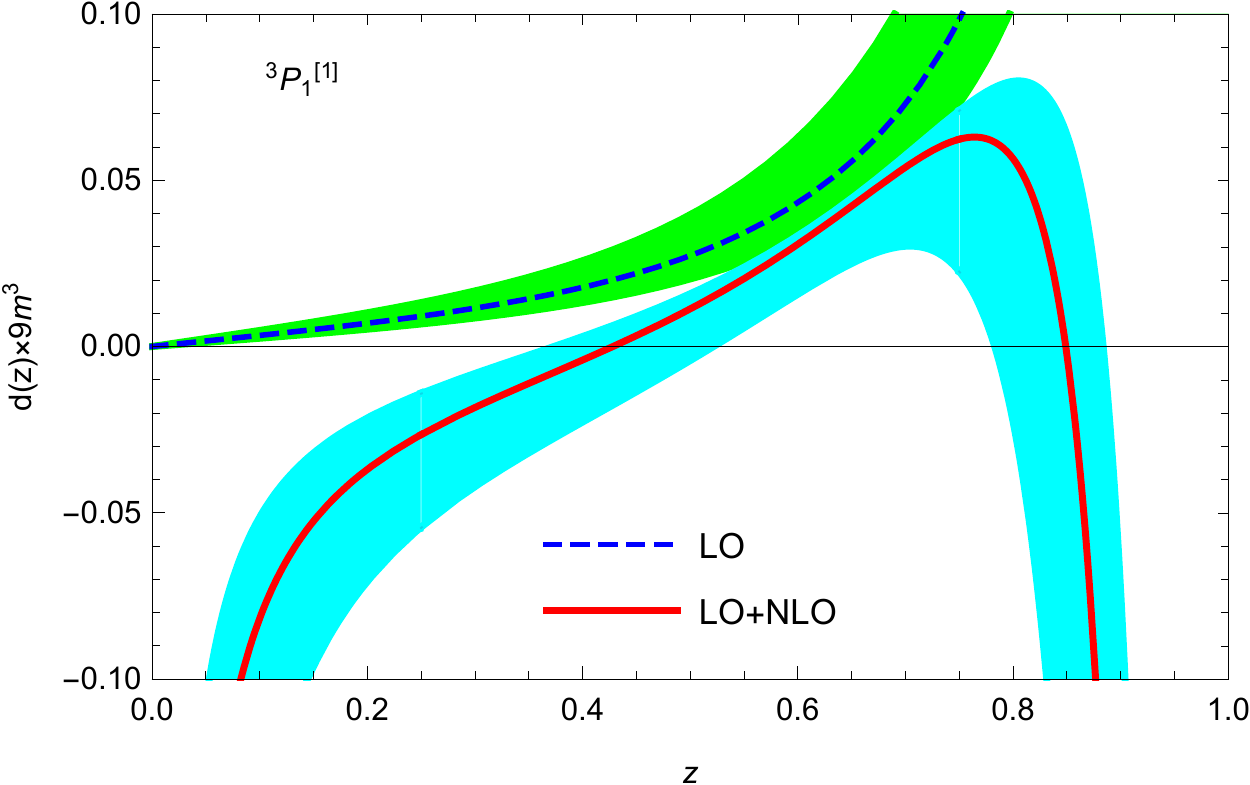}}
  \end{minipage}
\vfill
\vspace{5pt}
\begin{minipage}{0.5\textwidth}
  \centerline{$g\to b\bar{b} (\CScPb)+X $}
  \centerline{\includegraphics[width=1\textwidth]{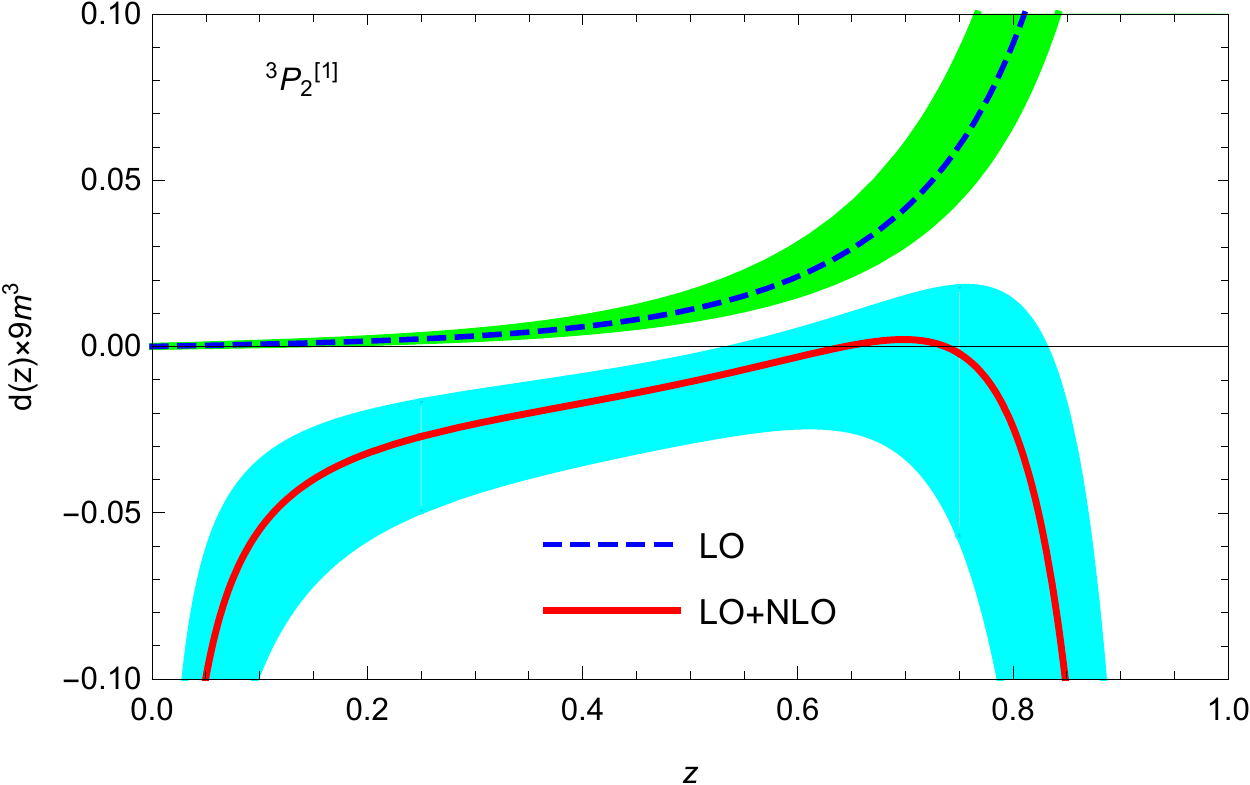}}
  \end{minipage}
\hfill
\begin{minipage}{0.5\textwidth}
  \centerline{$g\to b\bar{b} (\COcPj)+X $}
  \centerline{\includegraphics[width=1\textwidth]{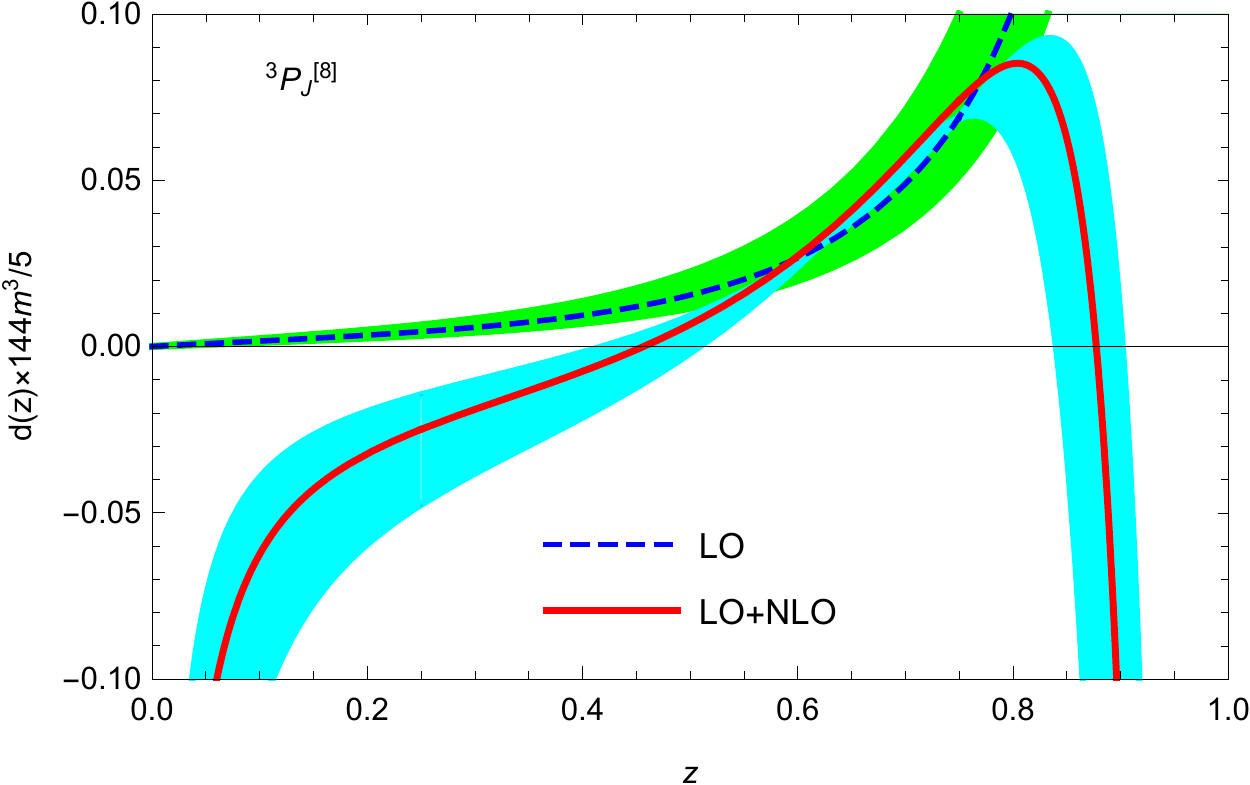}}
\end{minipage}
\caption{ SDCs of the fragmentation functions of $g\to b\bar{b} (\state{3}{P}{J}{1/8}))+X $ at LO and NLO. The dashed line is for $d_{\mathrm{LO}}(z)\times c^{[1/8]}$, and the solid line is for $(d_{\mathrm{LO}}(z)+ d_{\mathrm{NLO}}(z)) \times c^{[1/8]}$, with scale choices $\mu_r=\mu_0=\mu_\Lambda=2m_b$, $c^{[1]}=9m_b^5$ for color-singlet and $c^{[8]}=144m_b^5/5$ for color-octet. The bands are obtained by varying the renormalization scale $\mu_r$ by a factor of 2. }
\label{fig:res3pj18ren}
\end{figure}

To see the effects of the NLO corrections, we choose parameters with $m_b=4.75\gev$, $N_c=3$, $n_f=4$, and $\alpha_s (\mu_r=2m_b)=0.181 $.
In Figure \ref{fig:res3pj18ren}, we plot the curves of LO SDCs and LO+NLO SDCs for the $P$-wave FFs with $\mu_0=\mu_\Lambda = 2m_b$.
The overall factor $9m_b^5$ for color-singlet  and $144 m_b^5/5$ for color-octet comes from normalization factors at NLO.

The sensitivities of LO and LO+NLO FFs with respective to the renormalization scale $\mu_r$ are illustrated Fig.~\ref{fig:res3pj18ren} with bands corresponding to vary $\mu_r$ from $m_b$ to $4m_b$. We find that there are still large theoretical uncertainties after the NLO corrections.

With $\mu_r = \mu_0=\mu_\Lambda = 2m_b$, we also provide K-factors (the ratio of LO+NLO over LO) for some special values of $z$ in Table \ref{table:Kfactor} , where we find that the K-factors are negative for most values of $z$.
\begin{table}[htb!]
\begin{tabular*}{\textwidth}{@{\extracolsep{\fill}}*{1}{*{5}{c}}}
\hline \hline
$z$ & $K_0^{[1]}$ & $K_1^{[1]}$ & $K_2^{[1]}$ & $K_J^{[8]}$  \\
\hline
$0.10$ & $-5.746823334$ & $-24.60770220$ & $-79.84997431$ & $-38.52164437$ \\
$0.15$ & $-2.285338484$ & $-10.32801893$ & $-36.18197374$ & $-17.35843659$ \\
$0.20$ & $-2.644346438$ & $-5.288432534$ & $-19.87280192$ & $-9.481286949$ \\
$0.25$ & $-4.379318130$ & $-2.914573185$ & $-11.89225918$ & $-5.572752664$ \\
$0.30$ & $-7.198593861$ & $-1.593420613$ & $-7.390964744$ & $-3.288936921$ \\
$0.35$ & $-11.09174086$ & $-0.7748378276$ & $-4.639679071$ & $-1.813295182$ \\
$0.40$ & $-15.40618698$ & $-0.2304903644$ & $-2.878388543$ & $-0.8000547312$ \\
$0.45$ & $-17.98353020$ & $0.1472252728$ & $-1.722389454$ & $-0.08193653748$ \\
$0.50$ & $-16.79362533$ & $0.4131713893$ & $-0.9577973668$ & $0.4303450620$ \\
$0.55$ & $-13.06191610$ & $0.5961117069$ & $-0.4588481917$ & $0.7878655480$ \\
$0.60$ & $-9.270458907$ & $0.7101131977$ & $-0.1500652088$ & $1.020569545$ \\
$0.65$ & $-6.453740962$ & $0.7593385828$ & $0.01246796303$ & $1.144600900$ \\
$0.70$ & $-4.605946556$ & $0.7391584729$ & $0.05021547782$ & $1.164601499$ \\
$0.75$ & $-3.488264428$ & $0.6339314219$ & $-0.03511661296$ & $1.072131730$ \\
$0.80$ & $-2.912955012$ & $0.4091668085$ & $-0.2661028908$ & $0.8383434112$ \\
$0.85$ & $-2.799683132$ & $-0.01116815401$ & $-0.7094089942$ & $0.3920893805$ \\
$0.90$ & $-3.226904801$ & $-0.8227021717$ & $-1.548987722$ & $-0.4556362309$ \\
$0.95$ & $-4.779112355$ & $-2.756468771$ & $-3.481118189$ & $-2.413493587$ \\
\hline \hline
\end{tabular*}
\caption{K-factors at different values of $z$. $K_j^{[c]}$ denotes the K-factor of $g\to Q\bar{Q} (\state{3}{P}{j}{c})+X$ and the symbol $J$ means the sum of $j=0,1,2$. \label{table:Kfactor}}
\end{table}

As shown in Figure \ref{fig:res3pj18ren}, LO FFs are divergent at $z=1$.
% with NRQCD matching introduced in Sec. \ref{sec:match},
These divergences come from the $\delta(1-z)$ and $1/[1-z]_+$ terms in Eq.~\eqref{eq:sdclomu}.
Besides, the NLO FFs are negative and divergent at both $z=0$ and $z=1$. The leading behavior at $z\to 0$ is $1/z$, which means that the total fragmenting probability obtained by integrating the NLO FF over $z$ from $0$ to $1$ is divergent. But fortunately, physical cross sections are not sensitive to the behavior at $z\to 0$ after convolving FFs with partonic hard parts, which behave as $z^n$ in the small $z$ region with $n$ usually larger than $4$.
As shown in Eq.~\eqref{eq:sdcnlo}, the divergences at $z \to 1$ behavior as $\delta(1-z)$ and plus functions with leading power $[\ln^2(1-z)/(1-z) ]_+$, which comes from the expansion of $(1-z)^{-1-2\epsilon}\epsilon^{-2} $.

\begin{figure}
	\begin{minipage}{0.5\textwidth}
	  \centerline{$g\to b\bar{b} (\CScPz)+X $}
	  \centerline{\includegraphics[width=1\textwidth]{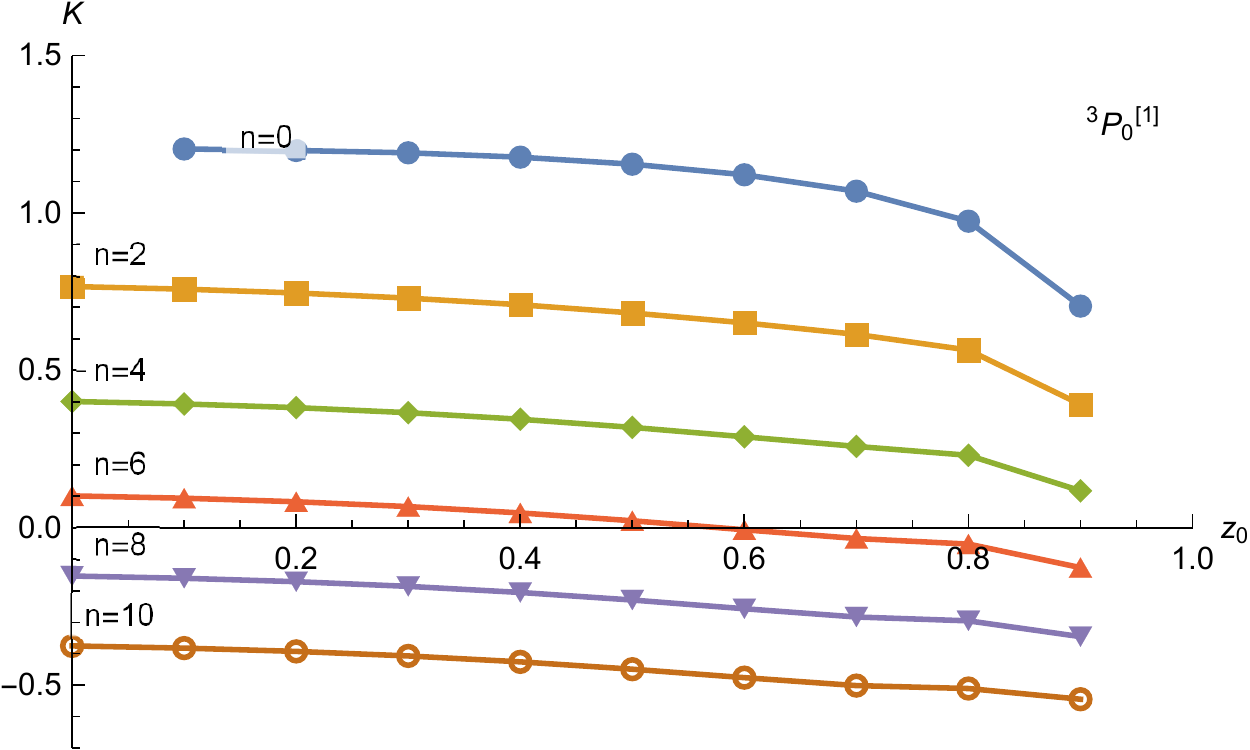}}
	\end{minipage}
	\hfill
	\begin{minipage}{0.5\textwidth}
	  \centerline{$g\to b\bar{b} (\CScPa)+X $}
	  \centerline{\includegraphics[width=1\textwidth]{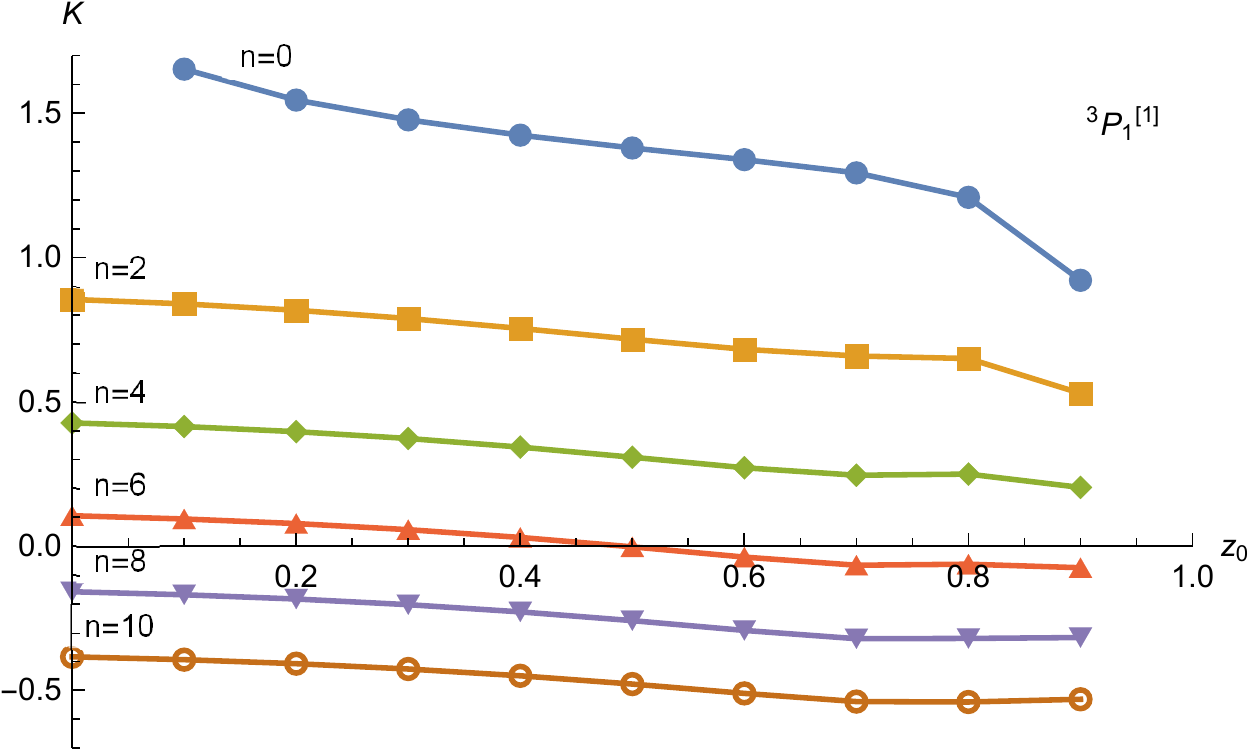}}
	\end{minipage}
	\vfill
	\vspace{5pt}
	\begin{minipage}{0.5\textwidth}
	  \centerline{$g\to b\bar{b} (\CScPb)+X $}
	  \centerline{\includegraphics[width=1\textwidth]{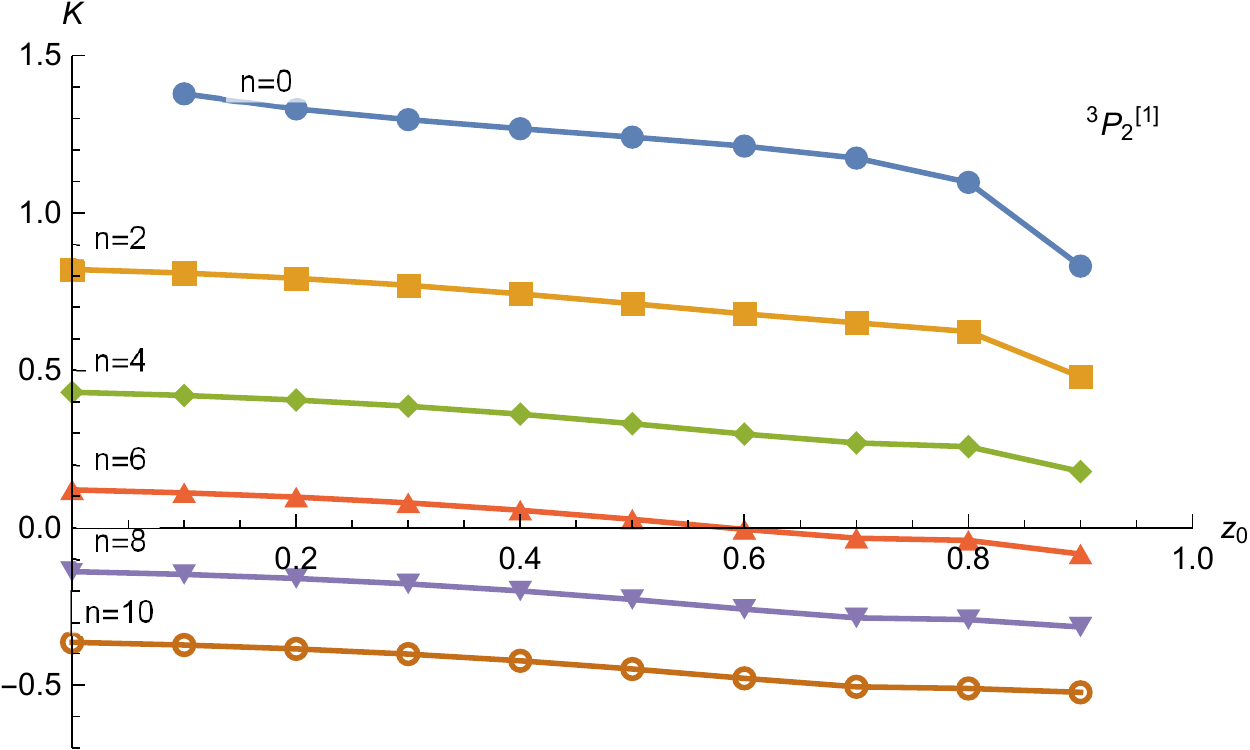}}
	\end{minipage}
	\hfill
	\begin{minipage}{0.5\textwidth}
	  \centerline{$g\to b\bar{b} (\COcPj)+X $}
	  \centerline{\includegraphics[width=1\textwidth]{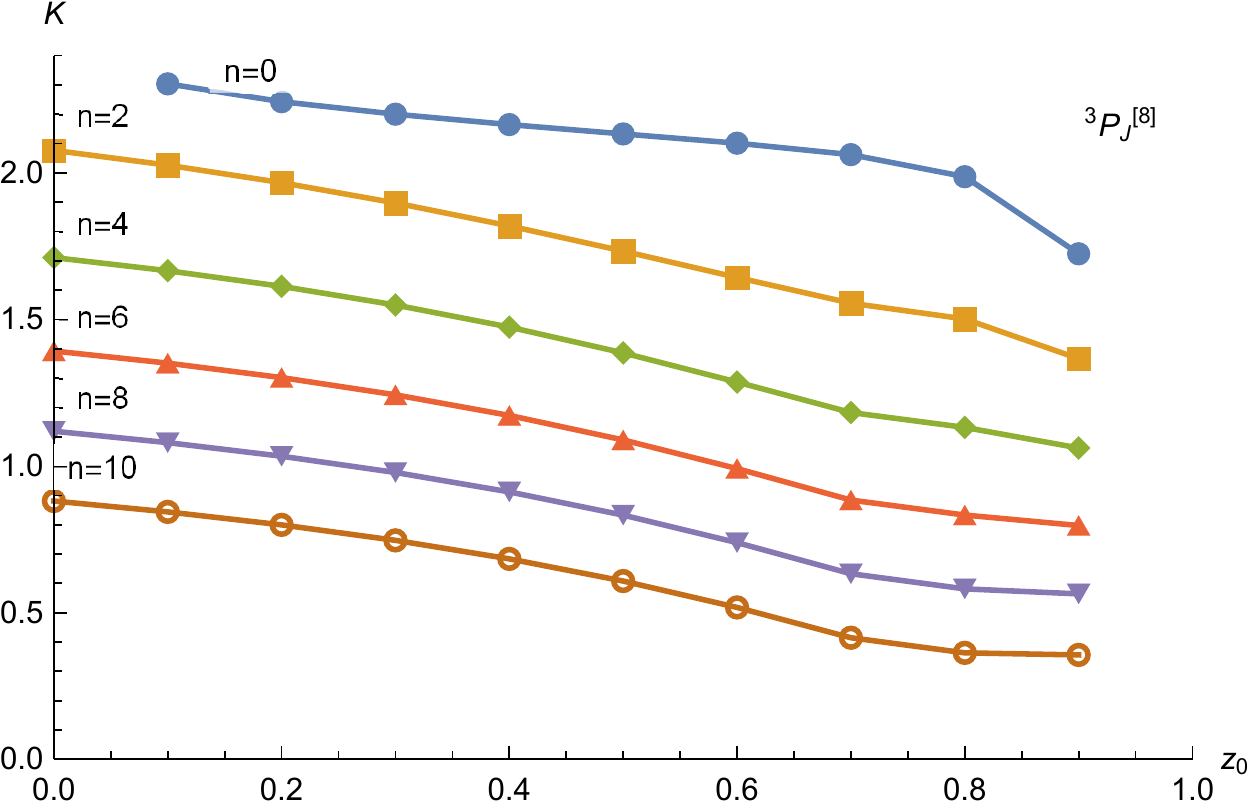}}
	\end{minipage}
	\caption{ The K-factors calculated by the integration results with Eq.~\eqref{eq:moments} for different $z_0$ and $n$. }
	\label{fig:intz0n}
\end{figure}

Obviously, the distribution of FF will not give us direct information about physical cross sections. To see the impact of the NLO results to the physical cross sections, we evaluate the $n$-th moments of the SDCs,
\begin{equation}\label{eq:moments}
	\int_{z_0}^1 \ud z \, z^n (d_{\mathrm{LO}}(z)+ d_{\mathrm{NLO}}(z))\times c^{[1/8]} \, ,
\end{equation}
where $c^{[1]}=9m_b^5$ and $c^{[8]}=144m_b^5/5$.
We plot the numerical results of integration K-factors with different $z_0$ and $n$, which are shown in Figure \ref{fig:intz0n}.
We find that for the same $n$, the K-factors are quite steady for different $z_0$, meaning that cross sections are dominated by FFs at large $z$.
Besides, the K-factors decrease with the increase of $n$. Especially, the color-singlet K-factors become negative for large $n$. And there may be a cancellation between the LO and the NLO FFs at about $7$-th moment for color-singlet quarkonium productions. Further more, it is clear that though the K-factors and values of FFs at each position $z=z_0$ are quite different for $j=0,1,2$ as shown in Table~\ref{table:Kfactor} and in Figure \ref{fig:res3pj18ren}, they are almost the same in the sense of integration as shown in Figure \ref{fig:intz0n}.

For the special case $z_0=0$, we can estimate the impact of NLO calculations on physical cross sections. The numerical results are shown in Table \ref{table:probability}.
\begin{table}[htb!]
	\begin{tabular*}{\textwidth}{@{\extracolsep{\fill}} c c c c c c c}
		\hline \hline
		state          			  & SDCs$*c^{[1/8]}$ 		  & $z^2$ & $z^4$ & $z^6$ & $z^8$ & $z^{10}$ \\
		\hline
		\multirow{3}{*}{$\CScPz$} & LO & $-0.128268$ & $-0.135763$ & $-0.141953$ & $-0.147253$ & $-0.151898$ \\
			& LO+NLO & $-0.0982734$ & $-0.0544311$ & $-0.0143470$ & $0.0226121$ & $0.0570568$ \\
			& K-factor & $0.766154$ & $0.400927$ & $0.101068$ & $-0.153559$ & $-0.375626$ \\
		\hline
		\multirow{3}{*}{$\CScPa$} & LO & $-0.0860886$ & $-0.102339$ & $-0.114018$ & $-0.123176$ & $-0.130722$ \\
			& LO+NLO & $-0.0735848$ & $-0.0436727$ & $-0.0119884$ & $0.0195102$ & $0.0501917$ \\
			& K-factor & $0.854756$ & $0.426745$ & $0.105145$ & $-0.158393$ & $-0.383957$ \\
		\hline
		\multirow{3}{*}{$\CScPb$} & LO & $-0.106291$ & $-0.118021$ & $-0.126989$ & $-0.134287$ & $-0.140456$ \\
			& LO+NLO & $-0.0872125$ & $-0.0507874$ & $-0.0152633$ & $0.0187392$ & $0.0511634$ \\
			& K-factor & $0.820505$ & $0.430326$ & $0.120194$ & $-0.139546$ & $-0.364266$ \\
		\hline
		\multirow{3}{*}{$\COcPj$} & LO & $-0.101999$ & $-0.114765$ & $-0.124328$ & $-0.132024$ & $-0.138483$ \\
			& LO+NLO & $-0.211825$ & $-0.196364$ & $-0.173108$ & $-0.147720$ & $-0.121978$ \\
			& K-factor & $2.07674$ & $1.71101$ & $1.39235$ & $1.11888$ & $0.880816$ \\
		\hline \hline
	\end{tabular*}
	\caption{Moments and K-factors of SDCs.  \label{table:probability}}
\end{table}
For large $n$, the influence of the behavior at $z\to 1$ becomes more and more prominent and the NLO correction becomes more and more significant.
In addition, the K-factors of color-octet process are much larger than those of color-singlet process.

The above discussion implies that resummation at large $z$ region may be crucial. This can be done within the framework of soft gluon factorization \cite{Ma:2017xno}, which we leave for future study.

%--------------------------------------------------------------------

\appendix
\section{Operator renormalization integrals} \label{sec:ori}

The operator convolution integration is shown in Eq.~\eqref{eq:operator} with the splitting functions shown in Eq.~\eqref{eq:splitting}.
The splitting functions can be divided into three parts, while the LO FF have two regions. Thus there are $6$ types of integrals,
\begin{align}
	& \int_z^1 \frac{\ud y}{y} \delta (1-y) (1-\frac{z}{y})^{-\epsilon} f_1\left(\frac{z}{y} \right)
		= \lambda^{-\epsilon} f_1(1-\lambda) \,,\\
	& \int_z^1 \frac{\ud y}{y} \delta (1-y) (1-\frac{z}{y})^{-1-2\epsilon} f_2\left(\frac{z}{y} \right)
		= \lambda^{-1-2\epsilon} f_2(1-\lambda) \,,\\
	\begin{split}
	& \int_z^1 \frac{\ud y}{y} g(y) (1-\frac{z}{y})^{-\epsilon} f_1\left(\frac{z}{y} \right) \\
	& = \lambda^{-\epsilon} \int_0^1 \ud y
		h(y,\epsilon)
		\frac{\lambda}{1-\lambda y}
		g\left( \frac{1-\lambda}{1-\lambda y} \right) f_1(1-\lambda y) \,,
	\end{split}\\
	\begin{split}
	& \int_z^1 \frac{\ud y}{y} g(y) (1-\frac{z}{y})^{-1-2\epsilon} f_2\left(\frac{z}{y} \right) \\
	&	= \lambda^{-1-2\epsilon}
		\Bigg[ -\frac{1}{2\epsilon} \lambda g(1-\lambda) f_2(1) \\
	& \phantom{=} + \int_0^1 \frac{\ud y }{y}
		h(y,2\epsilon)
		\left(
		\frac{\lambda}{1-\lambda y} g\left( \frac{1-\lambda}{1-\lambda y} \right) f_2(1-\lambda y)
		- \lambda g(1-\lambda) f_2(1)
		 \right)
		 \Bigg] \,,
	\end{split}\\
	\begin{split}
	& \int_z^1 \frac{\ud y}{y} \frac{1}{[1-y]_+} (1-\frac{z}{y})^{-\epsilon} f_1\left(\frac{z}{y} \right) \\
	& = \lambda^{-\epsilon} \ln \lambda \, f_1(1-\lambda)
	+ \lambda^{-\epsilon} \int_0^1 \frac{\ud y}{1-y}
		\left( h(y,\epsilon) f_1(1-\lambda y)
		- \frac{1-\lambda}{1-\lambda y} f_1(1-\lambda ) \right)  \,,
	\end{split}\\
	\begin{split}
	& \int_z^1 \frac{\ud y}{y} \frac{1}{[1-y]_+} (1-\frac{z}{y})^{-1-2\epsilon} f_2\left(\frac{z}{y} \right) \\
	& = \lambda^{-1-2\epsilon} \ln \lambda \, f_2(1-\lambda) \\
	& \phantom{=} + \lambda^{-1-2\epsilon}
		\Bigg[\left(-\frac{1}{2\epsilon} - H(-2\epsilon) -\ln (1-\lambda) \right) f_2(1) \\
	& \phantom{=} + \int_0^1 \frac{\ud y}{1-y}
		\left(
		h(y,2\epsilon) \frac{ f_2(1-\lambda y) - f_2(1) }{y}
		- \frac{1-\lambda}{1-\lambda y} \left( f_2(1-\lambda ) - f_2(1) \right)
		\right) \Bigg]\,,
	\end{split}
\end{align}
where $\lambda =1-z$, $g(y) = (1-2y)/y + y(1-y) $ comes from the splitting function, $H(-2\epsilon)$ is the Harmonic number which can be expanded in the limit of $\epsilon \to 0$ and
\begin{equation}
	h(y,\epsilon) = y^{-\epsilon} = \sum_i \frac{(-\epsilon)^i}{i!} \ln^i (y)
\end{equation}
is the Taylor series at $\epsilon \to 0$.
Based on these results, we find that  two more behaviors $(1-z)^{-\epsilon}\ln (1-z) $ and $(1-z)^{-1-2\epsilon}\ln (1-z) $ are added to the region at $z\to 1$.

%---------------------------------------------------------------------

%--------------------------------------------------------------------
% \begin{acknowledgments}
%--------------------------------------------------------------------

\acknowledgments

We thank Xiao Liu and Hao-Yu Liu for useful discussions.
The work is supported in part by the National Natural Science Foundation of China (Grants No. 11475005, No. 11075002, and No. 11875071), the National Key Basic Research Program of China (No. 2015CB856700), and High-performance Computing Platform of Peking University.

%--------------------------------------------------------------------
% \end{acknowledgments}
%--------------------------------------------------------------------

%--------------------------------------------------------------------
% \vspace{1.0 cm}

% \paragraph{Note added.} While this paper was being finalized, two related preprints appeared \cite{Artoisenet:2018dbs,Feng:2018ulg}. In Ref.~\cite{Artoisenet:2018dbs}, the authors calculated NLO corrections for FF of $g\to Q\bar{Q}(\state{1}{S}{0}{8})+X$ using FKS subtraction method; while in Ref.~\cite{Feng:2018ulg}, the authors calculated NLO corrections for FFs of $g\to Q\bar{Q}(\state{1}{S}{0}{1})+X$ and $g\to Q\bar{Q}(\state{1}{S}{0}{8})+X$ using sector decomposition. Our high-precision results agree with K-factors obtained in these two works within their estimated errors.

% $Note~ added.$ ---

%bibliography
\providecommand{\href}[2]{#2}\begingroup\raggedright\endgroup

% \bibliographystyle{jheppub}
% \bibliographystyle{utphysMa}
% \bibliography{article.jabref}
% \bibliography{article}

\begin{thebibliography}{10}

\bibitem{Bodwin:1994jh}
G.~T. Bodwin, E.~Braaten, and G.~P. Lepage, {\it {Rigorous QCD analysis of
  inclusive annihilation and production of heavy quarkonium}},
  \href{http://dx.doi.org/10.1103/PhysRevD.55.5853,
  10.1103/PhysRevD.51.1125}{{\em Phys. Rev.} {\bfseries D51} (1995) 1125--1171}
  [\href{http://arxiv.org/abs/hep-ph/9407339}{{\ttfamily hep-ph/9407339}}]
  [\href{http://inspirehep.net/search?p=find+Bodwin:1994jh}{{\ttfamily
  InSPIRE}}].
[Erratum: Phys. Rev.D55,5853(1997)].
%%CITATION = HEP-PH/9407339;%%.

\bibitem{Collins:1989gx}
J.~C. Collins, D.~E. Soper, and G.~Sterman, {\it {Factorization of Hard
  Processes in QCD}},
\href{http://dx.doi.org/10.1142/9789814503266_0001}{{\em Adv.Ser.Direct.High
  Energy Phys.} {\bfseries 5} (1988) 1--91}
  [\href{http://arxiv.org/abs/hep-ph/0409313}{{\ttfamily hep-ph/0409313}}]
  [\href{http://inspirehep.net/search?p=find+Collins:1989gx}{{\ttfamily
  InSPIRE}}].
%%CITATION = HEP-PH/0409313;%%.

\bibitem{Kang:2011mg}
Z.-B. Kang, J.-W. Qiu, and G.~Sterman, {\it {Heavy quarkonium production and
  polarization}},
\href{http://dx.doi.org/10.1103/PhysRevLett.108.102002}{{\em Phys.Rev.Lett.}
  {\bfseries 108} (2012) 102002}
  [\href{http://arxiv.org/abs/1109.1520}{{\ttfamily arXiv:1109.1520}}]
  [\href{http://inspirehep.net/search?p=find+Kang:2011mg}{{\ttfamily
  InSPIRE}}].
%%CITATION = ARXIV:1109.1520;%%.

\bibitem{Kang:2014tta}
Z.-B. Kang, Y.-Q. Ma, J.-W. Qiu, and G.~Sterman, {\it {Heavy quarkonium
  production at collider energies: Factorization and Evolution}},
\href{http://dx.doi.org/10.1103/PhysRevD.90.034006}{{\em Phys.Rev.} {\bfseries
  D90} (2014) 034006} [\href{http://arxiv.org/abs/1401.0923}{{\ttfamily
  arXiv:1401.0923}}]
  [\href{http://inspirehep.net/search?p=find+Kang:2014tta}{{\ttfamily
  InSPIRE}}].
%%CITATION = ARXIV:1401.0923;%%.

\bibitem{Kang:2014pya}
Z.-B. Kang, Y.-Q. Ma, J.-W. Qiu, and G.~Sterman, {\it {Heavy Quarkonium
  Production at Collider Energies: Partonic Cross Section and Polarization}},
\href{http://dx.doi.org/10.1103/PhysRevD.91.014030}{{\em Phys.Rev.} {\bfseries
  D91} (2015) 014030} [\href{http://arxiv.org/abs/1411.2456}{{\ttfamily
  arXiv:1411.2456}}]
  [\href{http://inspirehep.net/search?p=find+Kang:2014pya}{{\ttfamily
  InSPIRE}}].
%%CITATION = ARXIV:1411.2456;%%.

\bibitem{Fleming:2012wy}
S.~Fleming, A.~K. Leibovich, T.~Mehen, and I.~Z. Rothstein, {\it {The
  Systematics of Quarkonium Production at the LHC and Double Parton
  Fragmentation}},
\href{http://dx.doi.org/10.1103/PhysRevD.86.094012}{{\em Phys.Rev.} {\bfseries
  D86} (2012) 094012} [\href{http://arxiv.org/abs/1207.2578}{{\ttfamily
  arXiv:1207.2578}}]
  [\href{http://inspirehep.net/search?p=find+Fleming:2012wy}{{\ttfamily
  InSPIRE}}].
%%CITATION = ARXIV:1207.2578;%%.

\bibitem{Fleming:2013qu}
S.~Fleming, A.~K. Leibovich, T.~Mehen, and I.~Z. Rothstein, {\it {Anomalous
  dimensions of the double parton fragmentation functions}},
\href{http://dx.doi.org/10.1103/PhysRevD.87.074022}{{\em Phys.Rev.} {\bfseries
  D87} (2013) 074022} [\href{http://arxiv.org/abs/1301.3822}{{\ttfamily
  arXiv:1301.3822}}]
  [\href{http://inspirehep.net/search?p=find+Fleming:2013qu}{{\ttfamily
  InSPIRE}}].
%%CITATION = ARXIV:1301.3822;%%.

\bibitem{Gribov:1972ri}
V.~Gribov and L.~Lipatov, {\it {Deep inelastic e p scattering in perturbation
  theory}},
{\em Sov.J.Nucl.Phys.} {\bfseries 15} (1972) 438--450
  [\href{http://inspirehep.net/search?p=find+Gribov:1972ri}{{\ttfamily
  InSPIRE}}].
%%CITATION = SJNCA,15,438;%%.

\bibitem{Altarelli:1977zs}
G.~Altarelli and G.~Parisi, {\it {Asymptotic Freedom in Parton Language}},
\href{http://dx.doi.org/10.1016/0550-3213(77)90384-4}{{\em Nucl.Phys.}
  {\bfseries B126} (1977) 298}
  [\href{http://inspirehep.net/search?p=find+Altarelli:1977zs}{{\ttfamily
  InSPIRE}}].
%%CITATION = NUPHA,B126,298;%%.

\bibitem{Dokshitzer:1977sg}
Y.~L. Dokshitzer, {\it {Calculation of the Structure Functions for Deep
  Inelastic Scattering and $e^+ e^-$ Annihilation by Perturbation Theory in
  Quantum Chromodynamics.}},
{\em Sov.Phys.JETP} {\bfseries 46} (1977) 641--653
  [\href{http://inspirehep.net/search?p=find+Dokshitzer:1977sg}{{\ttfamily
  InSPIRE}}].
%%CITATION = SPHJA,46,641;%%.

\bibitem{Braaten:1993mp}
E.~Braaten, K.~Cheung, and T.~C. Yuan, {\it {$Z^0$ decay into charmonium via
  charm quark fragmentation}},
\href{http://dx.doi.org/10.1103/PhysRevD.48.4230}{{\em Phys.Rev.} {\bfseries
  D48} (1993) 4230--4235}
  [\href{http://arxiv.org/abs/hep-ph/9302307}{{\ttfamily hep-ph/9302307}}]
  [\href{http://inspirehep.net/search?p=find+Braaten:1993mp}{{\ttfamily
  InSPIRE}}].
%%CITATION = HEP-PH/9302307;%%.

\bibitem{Braaten:1993rw}
E.~Braaten and T.~C. Yuan, {\it {Gluon fragmentation into heavy quarkonium}},
\href{http://dx.doi.org/10.1103/PhysRevLett.71.1673}{{\em Phys.Rev.Lett.}
  {\bfseries 71} (1993) 1673--1676}
  [\href{http://arxiv.org/abs/hep-ph/9303205}{{\ttfamily hep-ph/9303205}}]
  [\href{http://inspirehep.net/search?p=find+Braaten:1993rw}{{\ttfamily
  InSPIRE}}].
%%CITATION = HEP-PH/9303205;%%.

\bibitem{Cho:1994gb}
P.~L. Cho, M.~B. Wise, and S.~P. Trivedi, {\it {Gluon fragmentation into
  polarized charmonium}},
\href{http://dx.doi.org/10.1103/PhysRevD.51.R2039}{{\em Phys. Rev.} {\bfseries
  D51} (1995) R2039--R2043}
  [\href{http://arxiv.org/abs/hep-ph/9408352}{{\ttfamily hep-ph/9408352}}]
  [\href{http://inspirehep.net/search?p=find+Cho:1994gb}{{\ttfamily InSPIRE}}].
%%CITATION = HEP-PH/9408352;%%.

\bibitem{Braaten:1994kd}
E.~Braaten and T.~C. Yuan, {\it {Gluon fragmentation into P wave heavy
  quarkonium}},
\href{http://dx.doi.org/10.1103/PhysRevD.50.3176}{{\em Phys.Rev.} {\bfseries
  D50} (1994) 3176--3180}
  [\href{http://arxiv.org/abs/hep-ph/9403401}{{\ttfamily hep-ph/9403401}}]
  [\href{http://inspirehep.net/search?p=find+Braaten:1994kd}{{\ttfamily
  InSPIRE}}].
%%CITATION = HEP-PH/9403401;%%.

\bibitem{Beneke:1995yb}
M.~Beneke and I.~Z. Rothstein, {\it {Psi-prime polarization as a test of color
  octet quarkonium production}},
  \href{http://dx.doi.org/10.1016/S0370-2693(96)80022-0,
  10.1016/0370-2693(96)00030-5}{{\em Phys. Lett.} {\bfseries B372} (1996)
  157--164} [\href{http://arxiv.org/abs/hep-ph/9509375}{{\ttfamily
  hep-ph/9509375}}]
  [\href{http://inspirehep.net/search?p=find+Beneke:1995yb}{{\ttfamily
  InSPIRE}}].
[Erratum: Phys. Lett.B389,769(1996)].
%%CITATION = HEP-PH/9509375;%%.

\bibitem{Ma:1995vi}
J.~P. Ma, {\it {Quark fragmentation into p wave triplet quarkonium}},
\href{http://dx.doi.org/10.1103/PhysRevD.53.1185}{{\em Phys. Rev.} {\bfseries
  D53} (1996) 1185--1190}
  [\href{http://arxiv.org/abs/hep-ph/9504263}{{\ttfamily hep-ph/9504263}}]
  [\href{http://inspirehep.net/search?p=find+Ma:1995vi}{{\ttfamily InSPIRE}}].
%%CITATION = HEP-PH/9504263;%%.

\bibitem{Braaten:1996rp}
E.~Braaten and Y.-Q. Chen, {\it {Dimensional regularization in quarkonium
  calculations}},
\href{http://dx.doi.org/10.1103/PhysRevD.55.2693}{{\em Phys.Rev.} {\bfseries
  D55} (1997) 2693--2707}
  [\href{http://arxiv.org/abs/hep-ph/9610401}{{\ttfamily hep-ph/9610401}}]
  [\href{http://inspirehep.net/search?p=find+Braaten:1996rp}{{\ttfamily
  InSPIRE}}].
%%CITATION = HEP-PH/9610401;%%.

\bibitem{Braaten:2000pc}
E.~Braaten and J.~Lee, {\it {Next-to-leading order calculation of the color
  octet 3S(1) gluon fragmentation function for heavy quarkonium}},
\href{http://dx.doi.org/10.1016/S0550-3213(00)00396-5}{{\em Nucl. Phys.}
  {\bfseries B586} (2000) 427--439}
  [\href{http://arxiv.org/abs/hep-ph/0004228}{{\ttfamily hep-ph/0004228}}]
  [\href{http://inspirehep.net/search?p=find+Braaten:2000pc}{{\ttfamily
  InSPIRE}}].
%%CITATION = HEP-PH/0004228;%%.

\bibitem{Hao:2009fa}
G.~Hao, Y.~Zuo, and C.-F. Qiao,
{\it {The Fragmentation Function of Gluon Splitting into P-wave Spin-singlet
  Heavy Quarkonium}},  [\href{http://arxiv.org/abs/0911.5539}{{\ttfamily
  arXiv:0911.5539}}]
  [\href{http://inspirehep.net/search?p=find+Hao:2009fa}{{\ttfamily InSPIRE}}].
%%CITATION = ARXIV:0911.5539;%%.

\bibitem{Jia:2012qx}
Y.~Jia, W.-L. Sang, and J.~Xu, {\it {Inclusive $h_c$ Production at $B$
  Factories}},
\href{http://dx.doi.org/10.1103/PhysRevD.86.074023}{{\em Phys. Rev.} {\bfseries
  D86} (2012) 074023} [\href{http://arxiv.org/abs/1206.5785}{{\ttfamily
  arXiv:1206.5785}}]
  [\href{http://inspirehep.net/search?p=find+Jia:2012qx}{{\ttfamily InSPIRE}}].
%%CITATION = ARXIV:1206.5785;%%.

\bibitem{Bodwin:2014bia}
G.~T. Bodwin, H.~S. Chung, U.-R. Kim, and J.~Lee, {\it {Quark fragmentation
  into spin-triplet $S$-wave quarkonium}},
\href{http://dx.doi.org/10.1103/PhysRevD.91.074013}{{\em Phys. Rev.} {\bfseries
  D91} (2015) 074013} [\href{http://arxiv.org/abs/1412.7106}{{\ttfamily
  arXiv:1412.7106}}]
  [\href{http://inspirehep.net/search?p=find+Bodwin:2014bia}{{\ttfamily
  InSPIRE}}].
%%CITATION = ARXIV:1412.7106;%%.

\bibitem{Ma:2013yla}
Y.-Q. Ma, J.-W. Qiu, and H.~Zhang, {\it {Heavy quarkonium fragmentation
  functions from a heavy quark pair. I. $S$ wave}},
\href{http://dx.doi.org/10.1103/PhysRevD.89.094029}{{\em Phys.Rev.} {\bfseries
  D89} (2014) 094029} [\href{http://arxiv.org/abs/1311.7078}{{\ttfamily
  arXiv:1311.7078}}]
  [\href{http://inspirehep.net/search?p=find+Ma:2013yla}{{\ttfamily InSPIRE}}].
%%CITATION = ARXIV:1311.7078;%%.

\bibitem{Ma:2015yka}
Y.-Q. Ma, J.-W. Qiu, and H.~Zhang, {\it {Fragmentation functions of polarized
  heavy quarkonium}},
\href{http://dx.doi.org/10.1007/JHEP06(2015)021}{{\em JHEP} {\bfseries 06}
  (2015) 021} [\href{http://arxiv.org/abs/1501.04556}{{\ttfamily
  arXiv:1501.04556}}]
  [\href{http://inspirehep.net/search?p=find+Ma:2015yka}{{\ttfamily InSPIRE}}].
%%CITATION = ARXIV:1501.04556;%%.

\bibitem{Zhang:2017xoj}
P.~Zhang, Y.-Q. Ma, Q.~Chen, and K.-T. Chao, {\it {Analytical calculation for
  the gluon fragmentation into spin-triplet S-wave quarkonium}},
\href{http://dx.doi.org/10.1103/PhysRevD.96.094016}{{\em Phys. Rev.} {\bfseries
  D96} (2017) 094016} [\href{http://arxiv.org/abs/1708.01129}{{\ttfamily
  arXiv:1708.01129}}]
  [\href{http://inspirehep.net/search?p=find+Zhang:2017xoj}{{\ttfamily
  InSPIRE}}].
%%CITATION = ARXIV:1708.01129;%%.

\bibitem{Braaten:1995cj}
E.~Braaten and T.~C. Yuan, {\it {Gluon fragmentation into spin triplet S wave
  quarkonium}},
\href{http://dx.doi.org/10.1103/PhysRevD.52.6627}{{\em Phys. Rev.} {\bfseries
  D52} (1995) 6627--6629}
  [\href{http://arxiv.org/abs/hep-ph/9507398}{{\ttfamily hep-ph/9507398}}]
  [\href{http://inspirehep.net/search?p=find+Braaten:1995cj}{{\ttfamily
  InSPIRE}}].
%%CITATION = HEP-PH/9507398;%%.

\bibitem{Bodwin:2003wh}
G.~T. Bodwin and J.~Lee, {\it {Relativistic corrections to gluon fragmentation
  into spin triplet S wave quarkonium}},
\href{http://dx.doi.org/10.1103/PhysRevD.69.054003}{{\em Phys. Rev.} {\bfseries
  D69} (2004) 054003} [\href{http://arxiv.org/abs/hep-ph/0308016}{{\ttfamily
  hep-ph/0308016}}]
  [\href{http://inspirehep.net/search?p=find+Bodwin:2003wh}{{\ttfamily
  InSPIRE}}].
%%CITATION = HEP-PH/0308016;%%.

\bibitem{Bodwin:2012xc}
G.~T. Bodwin, U.-R. Kim, and J.~Lee, {\it {Higher-order relativistic
  corrections to gluon fragmentation into spin-triplet S-wave quarkonium}},
\href{http://dx.doi.org/10.1007/JHEP11(2012)020}{{\em JHEP} {\bfseries 1211}
  (2012) 020} [\href{http://arxiv.org/abs/1208.5301}{{\ttfamily
  arXiv:1208.5301}}]
  [\href{http://inspirehep.net/search?p=find+Bodwin:2012xc}{{\ttfamily
  InSPIRE}}].
%%CITATION = ARXIV:1208.5301;%%.

\bibitem{Sun:2018yam}
Q.-F. Sun, Y.~Jia, X.~Liu, and R.~Zhu, {\it {Inclusive $h_c$ production and
  energy spectrum from $e^+e^-$ annihilation at a super $B$ factory}},
\href{http://dx.doi.org/10.1103/PhysRevD.98.014039}{{\em Phys. Rev.} {\bfseries
  D98} (2018) 014039} [\href{http://arxiv.org/abs/1801.10137}{{\ttfamily
  arXiv:1801.10137}}]
  [\href{http://inspirehep.net/search?p=find+Sun:2018yam}{{\ttfamily
  InSPIRE}}].
%%CITATION = ARXIV:1801.10137;%%.

\bibitem{Zhang:2018mlo}
P.~Zhang, C.-Y. Wang, X.~Liu, Y.-Q. Ma, C.~Meng, and K.-T. Chao, {\it
  {Semi-analytical calculation of gluon fragmentation into$^{1}$S$_{0}^{[1,8]}$
  quarkonia at next-to-leading order}},
\href{http://dx.doi.org/10.1007/JHEP04(2019)116}{{\em JHEP} {\bfseries 04}
  (2019) 116} [\href{http://arxiv.org/abs/1810.07656}{{\ttfamily
  arXiv:1810.07656}}]
  [\href{http://inspirehep.net/search?p=find+Zhang:2018mlo}{{\ttfamily
  InSPIRE}}].
%%CITATION = ARXIV:1810.07656;%%.

\bibitem{Artoisenet:2018dbs}
P.~Artoisenet and E.~Braaten, {\it {Gluon fragmentation into quarkonium at
  next-to-leading order using FKS subtraction}},
\href{http://dx.doi.org/10.1007/JHEP01(2019)227}{{\em JHEP} {\bfseries 01}
  (2019) 227} [\href{http://arxiv.org/abs/1810.02448}{{\ttfamily
  arXiv:1810.02448}}]
  [\href{http://inspirehep.net/search?p=find+Artoisenet:2018dbs}{{\ttfamily
  InSPIRE}}].
%%CITATION = ARXIV:1810.02448;%%.

\bibitem{Feng:2018ulg}
F.~Feng and Y.~Jia,
{\it {Next-to-leading-order QCD corrections to gluon fragmentation into
  ${}^1S_0^{(1,8)}$ quarkonia}},
  [\href{http://arxiv.org/abs/1810.04138}{{\ttfamily arXiv:1810.04138}}]
  [\href{http://inspirehep.net/search?p=find+Feng:2018ulg}{{\ttfamily
  InSPIRE}}].
%%CITATION = ARXIV:1810.04138;%%.

\bibitem{Nayak:2005rw}
G.~C. Nayak, J.-W. Qiu, and G.~F. Sterman, {\it {Fragmentation, factorization
  and infrared poles in heavy quarkonium production}},
  \href{http://dx.doi.org/10.1016/j.physletb.2005.03.031}{{\em Phys. Lett. B}
  {\bfseries 613} (2005) 45--51}
  [\href{http://arxiv.org/abs/hep-ph/0501235}{{\ttfamily hep-ph/0501235}}].

\bibitem{Nayak:2005rt}
G.~C. Nayak, J.-W. Qiu, and G.~Sterman, {\it {Fragmentation, NRQCD and NNLO
  factorization analysis in heavy quarkonium production}},
\href{http://dx.doi.org/10.1103/PhysRevD.72.114012}{{\em Phys.Rev.} {\bfseries
  D72} (2005) 114012} [\href{http://arxiv.org/abs/hep-ph/0509021}{{\ttfamily
  hep-ph/0509021}}]
  [\href{http://inspirehep.net/search?p=find+Nayak:2005rt}{{\ttfamily
  InSPIRE}}].
%%CITATION = HEP-PH/0509021;%%.

\bibitem{Nayak:2006fm}
G.~C. Nayak, J.-W. Qiu, and G.~F. Sterman, {\it {NRQCD Factorization and
  Velocity-dependence of NNLO Poles in Heavy Quarkonium Production}},
  \href{http://dx.doi.org/10.1103/PhysRevD.74.074007}{{\em Phys. Rev. D}
  {\bfseries 74} (2006) 074007}
  [\href{http://arxiv.org/abs/hep-ph/0608066}{{\ttfamily hep-ph/0608066}}].

\bibitem{Bodwin:2019bpf}
G.~T. Bodwin, H.~S. Chung, J.-H. Ee, U.-R. Kim, and J.~Lee, {\it {Covariant
  calculation of a two-loop test of nonrelativistic QCD factorization}},
  \href{http://dx.doi.org/10.1103/PhysRevD.101.096011}{{\em Phys. Rev. D}
  {\bfseries 101} (2020) 096011}
  [\href{http://arxiv.org/abs/1910.05497}{{\ttfamily arXiv:1910.05497}}].

\bibitem{Collins:1981uw}
J.~C. Collins and D.~E. Soper, {\it {Parton Distribution and Decay Functions}},
\href{http://dx.doi.org/10.1016/0550-3213(82)90021-9}{{\em Nucl.Phys.}
  {\bfseries B194} (1982) 445}
  [\href{http://inspirehep.net/search?p=find+Collins:1981uw}{{\ttfamily
  InSPIRE}}].
%%CITATION = NUPHA,B194,445;%%.

\bibitem{Kotikov:1990kg}
A.~V. Kotikov, {\it {Differential equations method: New technique for massive
  Feynman diagrams calculation}},
\href{http://dx.doi.org/10.1016/0370-2693(91)90413-K}{{\em Phys. Lett.}
  {\bfseries B254} (1991) 158--164}
  [\href{http://inspirehep.net/search?p=find+Kotikov:1990kg}{{\ttfamily
  InSPIRE}}].
%%CITATION = PHLTA,B254,158;%%.

\bibitem{Bern:1992em}
Z.~Bern, L.~J. Dixon, and D.~A. Kosower, {\it {Dimensionally regulated one loop
  integrals}},
\href{http://dx.doi.org/10.1016/0370-2693(93)90400-C}{{\em Phys.Lett.}
  {\bfseries B302} (1993) 299--308}
  [\href{http://arxiv.org/abs/hep-ph/9212308}{{\ttfamily hep-ph/9212308}}]
  [\href{http://inspirehep.net/search?p=find+Bern:1992em}{{\ttfamily
  InSPIRE}}].
%%CITATION = HEP-PH/9212308;%%.

\bibitem{Remiddi:1997ny}
E.~Remiddi, {\it {Differential equations for Feynman graph amplitudes}},
{\em Nuovo Cim.} {\bfseries A110} (1997) 1435--1452
  [\href{http://arxiv.org/abs/hep-th/9711188}{{\ttfamily hep-th/9711188}}]
  [\href{http://inspirehep.net/search?p=find+Remiddi:1997ny}{{\ttfamily
  InSPIRE}}].
%%CITATION = HEP-TH/9711188;%%.

\bibitem{Gehrmann:1999as}
T.~Gehrmann and E.~Remiddi, {\it {Differential equations for two loop four
  point functions}},
\href{http://dx.doi.org/10.1016/S0550-3213(00)00223-6}{{\em Nucl. Phys.}
  {\bfseries B580} (2000) 485--518}
  [\href{http://arxiv.org/abs/hep-ph/9912329}{{\ttfamily hep-ph/9912329}}]
  [\href{http://inspirehep.net/search?p=find+Gehrmann:1999as}{{\ttfamily
  InSPIRE}}].
%%CITATION = HEP-PH/9912329;%%.

\bibitem{Henn:2013pwa}
J.~M. Henn, {\it {Multiloop integrals in dimensional regularization made
  simple}},
\href{http://dx.doi.org/10.1103/PhysRevLett.110.251601}{{\em Phys. Rev. Lett.}
  {\bfseries 110} (2013) 251601}
  [\href{http://arxiv.org/abs/1304.1806}{{\ttfamily arXiv:1304.1806}}]
  [\href{http://inspirehep.net/search?p=find+Henn:2013pwa}{{\ttfamily
  InSPIRE}}].
%%CITATION = ARXIV:1304.1806;%%.

\bibitem{Lee:2014ioa}
R.~N. Lee, {\it {Reducing differential equations for multiloop master
  integrals}},
\href{http://dx.doi.org/10.1007/JHEP04(2015)108}{{\em JHEP} {\bfseries 04}
  (2015) 108} [\href{http://arxiv.org/abs/1411.0911}{{\ttfamily
  arXiv:1411.0911}}]
  [\href{http://inspirehep.net/search?p=find+Lee:2014ioa}{{\ttfamily
  InSPIRE}}].
%%CITATION = ARXIV:1411.0911;%%.

\bibitem{Adams:2017tga}
L.~Adams, E.~Chaubey, and S.~Weinzierl, {\it {Simplifying Differential
  Equations for Multiscale Feynman Integrals beyond Multiple Polylogarithms}},
\href{http://dx.doi.org/10.1103/PhysRevLett.118.141602}{{\em Phys. Rev. Lett.}
  {\bfseries 118} (2017) 141602}
  [\href{http://arxiv.org/abs/1702.04279}{{\ttfamily arXiv:1702.04279}}]
  [\href{http://inspirehep.net/search?p=find+Adams:2017tga}{{\ttfamily
  InSPIRE}}].
%%CITATION = ARXIV:1702.04279;%%.

\bibitem{Caffo:2008aw}
M.~Caffo, H.~Czyz, M.~Gunia, and E.~Remiddi, {\it {BOKASUN: A Fast and precise
  numerical program to calculate the Master Integrals of the two-loop sunrise
  diagrams}},
\href{http://dx.doi.org/10.1016/j.cpc.2008.10.011}{{\em Comput. Phys. Commun.}
  {\bfseries 180} (2009) 427--430}
  [\href{http://arxiv.org/abs/0807.1959}{{\ttfamily arXiv:0807.1959}}]
  [\href{http://inspirehep.net/search?p=find+Caffo:2008aw}{{\ttfamily
  InSPIRE}}].
%%CITATION = ARXIV:0807.1959;%%.

\bibitem{Czakon:2008zk}
M.~Czakon, {\it {Tops from Light Quarks: Full Mass Dependence at Two-Loops in
  QCD}},
\href{http://dx.doi.org/10.1016/j.physletb.2008.05.028}{{\em Phys. Lett.}
  {\bfseries B664} (2008) 307--314}
  [\href{http://arxiv.org/abs/0803.1400}{{\ttfamily arXiv:0803.1400}}]
  [\href{http://inspirehep.net/search?p=find+Czakon:2008zk}{{\ttfamily
  InSPIRE}}].
%%CITATION = ARXIV:0803.1400;%%.

\bibitem{Mueller:2015lrx}
R.~Mueller and D.~G. \"{O}zt\"{u}rk, {\it {On the computation of finite
  bottom-quark mass effects in Higgs boson production}},
\href{http://dx.doi.org/10.1007/JHEP08(2016)055}{{\em JHEP} {\bfseries 08}
  (2016) 055} [\href{http://arxiv.org/abs/1512.08570}{{\ttfamily
  arXiv:1512.08570}}]
  [\href{http://inspirehep.net/search?p=find+Mueller:2015lrx}{{\ttfamily
  InSPIRE}}].
%%CITATION = ARXIV:1512.08570;%%.

\bibitem{Lee:2017qql}
R.~N. Lee, A.~V. Smirnov, and V.~A. Smirnov, {\it {Solving differential
  equations for Feynman integrals by expansions near singular points}},
  \href{http://dx.doi.org/10.1007/JHEP03(2018)008}{{\em JHEP} {\bfseries 03}
  (2018) 008} [\href{http://arxiv.org/abs/1709.07525}{{\ttfamily
  arXiv:1709.07525}}].

\bibitem{Liu:2017jxz}
X.~Liu, Y.-Q. Ma, and C.-Y. Wang, {\it {A Systematic and Efficient Method to
  Compute Multi-loop Master Integrals}},
\href{http://dx.doi.org/10.1016/j.physletb.2018.02.026}{{\em Phys. Lett.}
  {\bfseries B779} (2018) 353--357}
  [\href{http://arxiv.org/abs/1711.09572}{{\ttfamily arXiv:1711.09572}}]
  [\href{http://inspirehep.net/search?p=find+Liu:2017jxz}{{\ttfamily
  InSPIRE}}].
%%CITATION = ARXIV:1711.09572;%%.

\bibitem{Liu:2020kpc}
X.~Liu, Y.-Q. Ma, W.~Tao, and P.~Zhang, {\it {Calculation of Feynman loop
  integration and phase-space integration via auxiliary mass flow}},
  [\href{http://arxiv.org/abs/2009.07987}{{\ttfamily arXiv:2009.07987}}].

\bibitem{ferguson1999analysis}
H.~R.~P. {Ferguson}, D.~H. {Bailey}, and S.~{Arno}, {\it {Analysis of PSLQ, an
  integer relation finding algorithm}},  {\em Mathematics of Computation}
  {\bfseries 68} (Jan., 1999) 351--369.

\bibitem{Bailey:1999nv}
D.~H. Bailey and D.~J. Broadhurst, {\it {Parallel integer relation detection:
  Techniques and applications}},
  \href{http://dx.doi.org/10.1090/S0025-5718-00-01278-3}{{\em Math. Comput.}
  {\bfseries 70} (2001) 1719--1736}
  [\href{http://arxiv.org/abs/math/9905048}{{\ttfamily math/9905048}}].

\bibitem{Ma:2017xno}
Y.-Q. Ma and K.-T. Chao, {\it {New factorization theory for heavy quarkonium
  production and decay}},
  \href{http://dx.doi.org/10.1103/PhysRevD.100.094007}{{\em Phys. Rev. D}
  {\bfseries 100} (2019) 094007}
  [\href{http://arxiv.org/abs/1703.08402}{{\ttfamily arXiv:1703.08402}}].

\end{thebibliography}
%---------------------------------------------------------------------

\end{document}